\documentclass[10pt,letterpaper]{article}
\usepackage[top=0.85in,left=2.75in,footskip=0.75in,marginparwidth=2in]{geometry}

\usepackage[utf8]{inputenc}

\usepackage{cite}

\usepackage{nameref, hyperref}
\hypersetup{
    colorlinks,
    linkcolor={black!50!black},
    citecolor={black!50!black},
    urlcolor={black!80!black}
}

\usepackage[right]{lineno}

\usepackage{microtype}
\DisableLigatures[f]{encoding = *, family = * }

\raggedright
\usepackage{changepage}

\usepackage[aboveskip=1pt,labelfont=bf,labelsep=period,singlelinecheck=off]{caption}

\usepackage{booktabs, multirow} 
\usepackage{changepage} 
\usepackage[flushleft]{threeparttable} 

\usepackage{lastpage,fancyhdr,graphicx}
\usepackage{epstopdf}
\fancyheadoffset[L]{2.25in}
\fancyfootoffset[L]{2.25in}

\usepackage{graphicx}

\usepackage{sidecap}

\usepackage{wrapfig}
\usepackage[pscoord]{eso-pic}
\usepackage[fulladjust]{marginnote}
\reversemarginpar

\begin{document}
\vspace*{0.35in}

\begin{flushleft}
{\Large
\textbf\newline{Evaluating Amazon Effects and the Limited Impact of COVID-19 With Purchases Crowdsourced from US Consumers}
}
\newline
\\
Alex Berke\textsuperscript{1,*},
Dana Calacci\textsuperscript{2},
Alex (Sandy) Pentland\textsuperscript{1,3},
Kent Larson\textsuperscript{1}
\\
\bigskip
\bf{1} Media Lab, Massachusetts Institute of Technology
\\
\bf{2} Penn State University
\\
\bf{3} HAI, Stanford University
\\
\bigskip
*aberke@mit.edu

\end{flushleft}

\section*{Abstract}
We leverage a recently published dataset of Amazon purchase histories, crowdsourced from thousands of US consumers, to study how online purchasing behaviors have changed over time, how changes vary across demographic groups, the impact of the COVID-19 pandemic, and relationships between online and offline retail. This work provides a case study in how consumer-level purchases data can reveal purchasing behaviors and trends beyond those available from aggregate metrics. For example, in addition to analyzing spending behavior, we develop new metrics to quantify changes in consumers' online purchase frequency and the diversity of products purchased, to better reflect the growing ubiquity and dominance of online retail. Between 2018 and 2022 these consumer-level metrics grew on average by more than 85\%, peaking in 2021. We find a steady upward trend in individuals' online purchasing prior to COVID-19, with a significant increase in the first year of COVID, but without a lasting effect. Purchasing behaviors in 2022 were no greater than the result of the pre-pandemic trend. We also find changes in purchasing significantly differ by demographics, with different responses to the pandemic. We further use the consumer-level data to provide evidence of substitution effects between online and offline retail in sectors where Amazon heavily invested: books, shoes, and grocery. Prior to COVID we find year-to-year changes in the number of consumers making online purchases for books and shoes negatively correlated with changes in employment at local bookstores and shoe stores. During COVID we find online grocery purchasing negatively correlated with in-store grocery visits. This work demonstrates how crowdsourced, open purchases data can enable economic insights that may otherwise only be available to private firms.

\section{Introduction}

The rise of e-commerce, led by Amazon, is transforming consumer behavior and retail markets. According to the US Census Bureau, e-commerce sales accounted for 9.4\% of total US retail in 2018, growing to 14\% by the end of 2022 ~\cite{censusEcommerce}. Amazon commands an estimated 38\% of this online market share, dwarfing its closest competition with less than 7\%~\cite{emarketerRanks2023}. This dominance has led to the popular use of the term ``Amazon effects" to describe changes ranging from new online consumer habits~\cite{mamaghani2020} and the resulting logistics issues for retailers~\cite{daugherty2019} to the shifting relationships between online and offline retail~\cite{AmazonEffectsInvestopedia}. The US Census Bureau provides quarterly estimates on e-commerce market share, but the data are highly aggregated, provided voluntarily by an undisclosed sample of firms~\cite{censusEcommerce}. Detailed data needed to understand the transformative ``Amazon effects" remain largely within companies like Amazon, inaccessible to researchers and the public. 

We address this public knowledge gap by analyzing a unique dataset of Amazon purchase histories crowdsourced from thousands of US consumers~\cite{ourSciDataPaper}. We leverage the disaggregated nature of this dataset to produce consumer-level metrics, providing new insights into how online retail is impacting consumers' purchasing behaviors and the ``Amazon effects” impacting offline retail. This work presents a case study in how disaggregated purchases data can reveal trends beyond those available from aggregate metrics, which future work can build upon.

For example, while consumer behavior has traditionally been tracked by expenditure, this risks conflating changes in consumer behavior and prices. We introduce metrics that quantify changes in consumers' online purchase frequency and product diversity, which we argue better reflect how the increasing dominance and convenience of online retail are changing consumer behaviors. We use these metrics to analyze relationships between consumer demographics and e-commerce growth, and uncover the limited effects of COVID-19 on pre-pandemic growth trends. 

Previous research has investigated how online retail has led to more dynamic pricing~\cite{chen2016}, finding prices are updated more frequently for goods available on Amazon~\cite{cavallo2018}, as well as more uniform pricing across store locations~\cite{ater2018,cavallo2018}. Research has also shown the effects Amazon can have on local retailers and employment. A study using data from 2010 to 2016 found the rollout of Amazon fulfillment centers reduced sales and employment at geographically proximate retail stores~\cite{chava2023}. Earlier work using data from Amazon's online books marketplace has found substitution effects between used and new books~\cite{ghose2006}, and local bookstore openings and online sales~\cite{forman2009}. We add to this literature with analysis beyond just books; we analyze three retail sectors in which Amazon substantially invested: books~\cite{amazon1998}, shoes~\cite{zapposSEC2009}, and grocery~\cite{sec2017}. We demonstrate substitution effects between Amazon and brick-and-mortar retail for each of these sectors, contributing to a larger discussion of the Amazon effect on local stores.

Much more e-commerce research emerged from the COVID-19 pandemic, with speculation that pandemic related shocks, such as store closures and stay-at-home orders, could accelerate the growth of online retail~\cite{pantano2020,ratchford2022,roggeveen2020}. However, much of this research was limited to the pandemic period without studying long-term effects. For example, the primary report released by the US Census Bureau on pandemic economic impacts showed substantial increases in e-commerce sales, but the analysis was limited to 2020 changes~\cite{censusbureauPandemicEconImpact}.  This work provides follow-up analysis to help evaluate the lasting impacts of COVID disruptions, leveraging longitudinal purchases data spanning the pre-pandemic (2018) to post-pandemic (2022) periods.  Contrary to speculation, we find COVID had a limited impact on trends in Amazon users' purchasing behaviors: COVID-19 significantly increased online purchasing temporarily, but purchasing behaviors in 2022 were no greater than the result of the pre-pandemic trend. We further show differences in purchasing behaviors between demographic groups and how COVID impacted these groups differently.

Overall, our analyses help reveal nuance in the relationship between e-commerce and the pandemic, its diverse growth across consumer groups, and substitution effects impacting offline retail.

\section{Materials and methods}

\subsection{Purchases and demographics data}
Our analyses use an open dataset containing purchase histories and user demographics collected from N=5027 US Amazon.com users~\cite{ourSciDataPaper}. The data were crowdsourced via an online survey and published with users' informed consent. Each user in the data contributed an export of their purchase histories from January 2018 to November 2022. The data collection process was previously described in~\cite{ourCSCWpaper}, with the dataset detailed in~\cite{ourSciDataPaper}. Here we briefly describe aspects of the data pertinent to our analyses.

\subsubsection{Amazon purchases}
Each row in the purchases dataset represents a purchase by a particular user and includes an order date, product code (ASIN/ISBN), product title, per-unit price and quantity, state the item was shipped to, and a category assigned by Amazon. Purchases are linked to a single user and their demographics data via a response ID. To better illustrate the dataset, we provide an example set of rows in SI Table~\ref{tab:s1}.

\subsubsection{User demographics}
Users who contributed their purchase histories also reported their demographics through a survey. The demographics include gender, age group, household income group, race and ethnicity, household size, and their US state of residence in 2021. Demographics used in the analyses are summarized in SI Tables~\ref{tab:s2}-\ref{tab:s4}. Users who reported gender outside the male/female binary or did not disclose income are dropped from analyses that use the sex and income variables. The survey allowed users to identify as multiple races; our analysis that uses race includes users in all race groups they selected, allowing users to be in multiple groups (see Fig~\ref{fig:purchase_freq_diffs_by_demos}).

\subsection{Metrics for estimating changes in purchasing behavior}

\begin{figure}
\begin{adjustwidth}{-1.9in}{0in}
    \begin{flushright}
    \includegraphics[width=\linewidth]{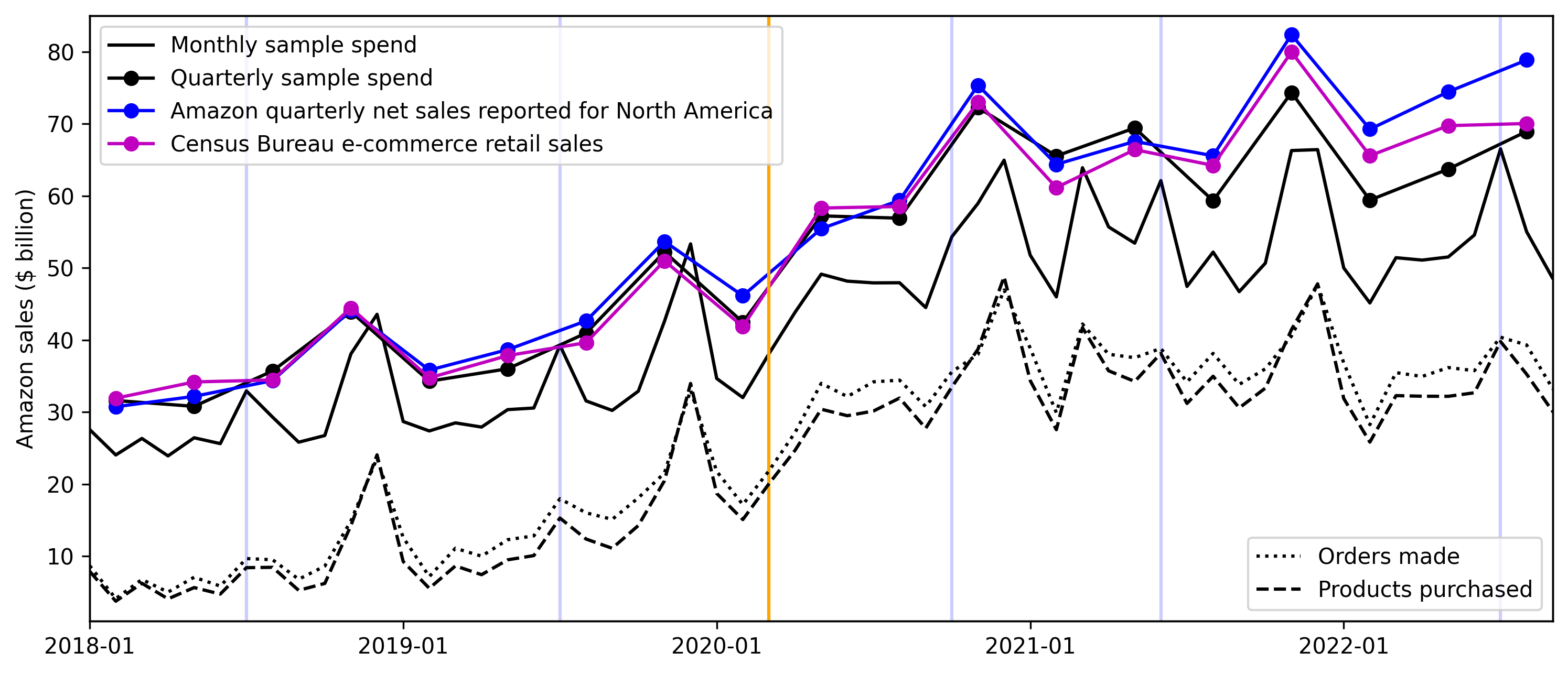}
    \caption{Quarterly Amazon net sales (North America segment) reported for investor relations and census e-commerce sales data, compared to metrics computed from our sample. Vertical blue lines indicate months Amazon Prime Day occurred. The orange line indicates March 2020, when COVID-19 had a major impact on US consumption. The sample metrics are scaled and shifted for legibility and should not be interpreted numerically.}
    \label{fig:amzn_sample_spend_vs_net_sales}
    \end{flushright}
\end{adjustwidth}
\end{figure}

Fig \ref{fig:amzn_sample_spend_vs_net_sales} provides a visual overview of the data and metrics we use to analyze changes in consumer purchasing behaviors, showing the following metrics computed from our sample data: total expenditure, the number of distinct products purchased and the number of orders made (measured as per person purchase days).  These metrics are shown alongside Amazon net sales data (North America segment), reported for investor relations~\cite{AmazonIRQuarterly}, and e-commerce retail sales data from the US Census Bureau~\cite{ECOMNSA2024}, which are both reported on a quarterly basis. We use these quarterly sales data to validate our metrics and then use our metrics to help reveal details the quarterly sales data lack, including changes within quarters and how changes are driven by different consumer groups. When comparing our sample's quarterly expenditure to the Amazon net sales and census e-commerce sales data, there is a Pearson correlation of $r=0.976$ ($p<0.001$) and $r=0.982$ ($p<0.001$), respectively. Fig \ref{fig:amzn_sample_spend_vs_net_sales} also shows how our sample's expenditure grew less quickly than Amazon sales in later quarters, which is expected given our sample is limited to a consistent set of users while Amazon's user population grew over time. Vertical blue lines in Fig \ref{fig:amzn_sample_spend_vs_net_sales} indicate months when Amazon's major sales event, Prime Day, occurred~\cite{amazonstaff2019}. We include these indicators throughout the results because Prime Day significantly increased the monthly metrics (see SI Table~\ref{tab:s5}).

For the following analyses we use the metrics shown in Fig~\ref{fig:amzn_sample_spend_vs_net_sales} computed as panel data.

\subsubsection{Preprocessing and panel data}
To produce the panel data used to estimate consumer behavior metrics, we first restrict the data to purchases made before November 2022. We filter out gift card purchases and restrict the data to users with purchases in both 2018 and 2022. We then compute panel data with a row for each user-month, for each month in the span of 2018-01 to 2022-10. In order to estimate regressions that assume linear trends, we add a column variable, $t$, to the panel, numbering the months starting at $t=0$ for 2018-01. Each row of panel data includes the monthly metrics for the corresponding user and month.

\subsubsection{Metrics}
We compute the following monthly metrics for each user-month in the panel data.

\noindent\textbf{Spend} is the total spend summed across users' purchases in USD.

\noindent\textbf{Distinct products} is the number of unique product IDs the user purchased.

\noindent\textbf{Purchase days} is the number of unique days the user made a purchase.

\subsection{Event study evaluating trends in purchasing behavior and the impact of COVID-19}

We produce a graphical event study in order to evaluate trends in consumers' purchasing behaviors and the impact of COVID. To do this, we first estimate the following OLS regression model (Eq~\ref{eq:graphical_event_study_1}) using the panel data, which measures the average change in purchase days per month, per user, controlling for sex, age, and income demographics.

\begin{equation}
    \label{eq:graphical_event_study_1}
    purchaseDays_{i,t} = a_i + \sum_t B_t \cdot T_{i,t}  + \beta \cdot X_i + e_i	
\end{equation}

The $a_i$ terms capture fixed effects for each user, $X_i$ represents a vector of consumer variables including sex (male/female), age group, and income group; $T_{i,t}$ is an indicator variable for each time period, $t$. A coefficient, $B_t$, is estimated for each $t$. We plot and further analyze the estimated $B_t$ coefficients (Fig~\ref{fig:event_study_purchase_days}). 

We use the $B_t$ coefficients estimated for months 2018-01 to 2020-02 via Eq~\ref{eq:graphical_event_study_1} in order to estimate a trend line prior to COVID. We do this via the following regression (Eq~\ref{eq:graphical_event_study_2}). Tabular results are in SI Table~\ref{tab:s10}.

\begin{equation}
    \label{eq:graphical_event_study_2}
B_t = intercept + \beta \cdot t	
\end{equation}

We reproduce this analysis using $distinctProducts$ and spend as the dependent variable in Eq~\ref{eq:graphical_event_study_1}, instead of $purchaseDays$.

We also test for whether there is a statistically significant increase in monthly purchasing via the following event study OLS model (Eq~\ref{eq:event_study_3}):

\begin{equation}
    \label{eq:event_study_3}
    purchaseDays_{i,t} = a_i + \beta_1 \cdot t  +  \beta_2 \cdot postCOVID_t + \beta_3 \cdot X_i  + \sum_m \beta_m \cdot month_t + e_i	  
\end{equation}

Variable $a_i$ captures the fixed effects for each user. We include an indicator variable for each month, $m$, to capture seasonality. $X_i$ represents a vector of consumer variables including sex (male/female), age group, and income group.  The variable of interest is $postCOVID$, which is 1 if 2020-03 or later; 0 otherwise. We limit this event study to dates before 2021-03 in order to compare the period before COVID (2018-01 to 2020-02) to the first year impacted by COVID (2020-03 to 2021-02). Regression results are shown in SI Table~\ref{tab:s11}. The positive coefficient for the $postCOVID$ variable ($\beta=0.5578$; $p<0.001$) indicates the additional purchase days in the $postCOVID$ period are significantly higher than the trend.

\subsection{Analyzing relationships between consumer demographics and purchasing}

We evaluate relationships between consumer demographics and purchase frequency with the following OLS regression (Eq~\ref{eq:demos_purchase_freq}).

\begin{equation}
    \label{eq:demos_purchase_freq}
    y_i = intercept + \beta \cdot X_i 
\end{equation}

The independent variables are consumer variables represented by vector $X_i$, specific to each user, $i$. The $X_i$ variables (shown in Fig~\ref{fig:purchase_freq_diffs_by_demos} and Table~\ref{tab:s12}) include sex, age group, income group, race and ethnicity, and household size. $X_i$ further controls for geography by including US state of residence. We run four separate OLS regressions, each using the same independent variables and only differing in the dependent variable, $y_i$. For (1) 2018 and (2) 2022 results, $y$ = purchase days per month, computed as the median over all months in the year, excluding November and December.  We exclude November and December because the data go up to November 2022 and the 2018 and 2022 results are meant to be comparable. We also evaluate changes over time: We (3) evaluate changes from 2018-2022, where $y$ = percent change in total purchase days in 2018 versus 2022, and we (4) evaluate changes over the first year of COVID, where $y$ = percent change in total purchase days from the year prior to COVID (2019-03 to 2020-02) to the period spanning the first year of COVID-19 (2020-03 to 2021-02).

\subsection{Analyzing online purchases versus retail employment prior to COVID}

To investigate locale-specific substitution effects prior to COVID, we test whether there is a negative correlation between changes in online purchases for category X and change in local employment at retail establishments of category X.  We test this hypothesis for year-to-year changes in the years prior to COVID, 2018 to 2019, for X=books and shoes.  To do this, we use yearly state level employment data for book stores and shoe stores retail establishments (NAICS codes 451211 and 448210, respectively) from the US Census Bureau Statistics of US Businesses (SUSB) tables~\cite{SUSB}. 

We do this analysis for books and shoes because these products belong to well defined retail sectors where (a) Amazon substantially invested and (b) there are available employment data from the Census Bureau (i.e. there are corresponding NAICS codes). We limit the analysis to 2018 and 2019 because the purchases data starts in 2018, employment (SUSB) data were only available up to 2021 at the time of analysis, and the years 2020 and 2021 were impacted by COVID.

When computing changes in the Amazon purchases data by state, we use the shipping address state associated with each purchase.  We restrict the analysis to users with purchases in 2018 and to states where at least n=50 users made purchases. 45 states (including DC) meet this threshold. This excludes AK, MT, ND, PR, SD, VT, WY, and results in N=4211 users across these 45 states. 

We compute the correlation between (1) the percent change in the 2018 to 2019 employment for X=book/shoe stores in each state and (2) the percent change in the number of users making purchases for books/shoes in each state. For computing (1) the percent change in employment, we first normalize the number of employees for type X establishments in each state by the state population, specific to the year~\cite{censusPop2021}, in order to account for employment changes due to population changes. We then compute percent change as: ($employees_{2019} - employees_{2018}) / employees_{2018}$. For (2) we first compute the number of distinct purchasers for category X in each state in each year as the mean of a repeated random sampling process. We randomly draw, without replacement, n=2500 users, compute the number of distinct users buying X in each state for each year, and repeat the process 1000 times.  The resulting negative correlation is statistically significant at the $p<0.05$ level. Pearson $r=-0.344$ ($p=0.021$) and $r=-0.315$ ($p=0.035$) for books and shoes, respectively.

\subsection{Analyzing online grocery purchases versus in-store shopping during COVID}
A number of technology companies that collect location data from mobile phones produced mobility datasets for researchers to study the COVID-19 pandemic~\cite{berkeMobilePhonesCovid2024}.  We use the Google COVID-19 Community Mobility Reports data~\cite{googlellc2020} for ``Grocery and pharmacy" produced for each US state, which measures how in-person visits and length of stay at these places changed compared to a baseline. (The baseline was the 5-week period of January 3 to February 6, 2020.) The data from Google is a daily metric, which we aggregate to a monthly metric by taking the monthly mean. We compare the monthly mobility metric to the monthly number of distinct users who made Amazon grocery purchases in the corresponding US state. Grocery purchases were identified using the product categories listed in the SI.

We test for a negative correlation between the mobility metric, which represents presence at grocery stories, and the number of users purchasing groceries online. We test this relationship for all months where Google mobility data are available, 2020-02 to 2022-01, excluding the December months (2020-12, 2021-12) where the holidays increase shopping overall, both online and offline. We test the top 3 US states, ordered by the average number of monthly Amazon users purchasing groceries: CA, TX, NY. We note that when ordering by population, the top US states are (1) CA, (2) TX, (3), FL, (4) NY~\cite{censusPopStates2022}.  We use the number of purchasers rather than the US population (which excludes FL) partly because states like FL have a seasonal population, where a population influx can contribute to both in-store purchases (captured by the mobility index) and online purchases. We find the expected negative correlation with Pearson $r=-0.544$ ($p=0.007$), $r=-0.431$ ($p=0.040$), and $r=-0.513$ ($p=0.012$), for CA, TX, NY, respectively.

\section{Results}

\subsection{Changes in consumer behavior}
\begin{figure}[ht]
\begin{adjustwidth}{-0.575in}{0in}
    \centering
    \includegraphics[width=\linewidth]{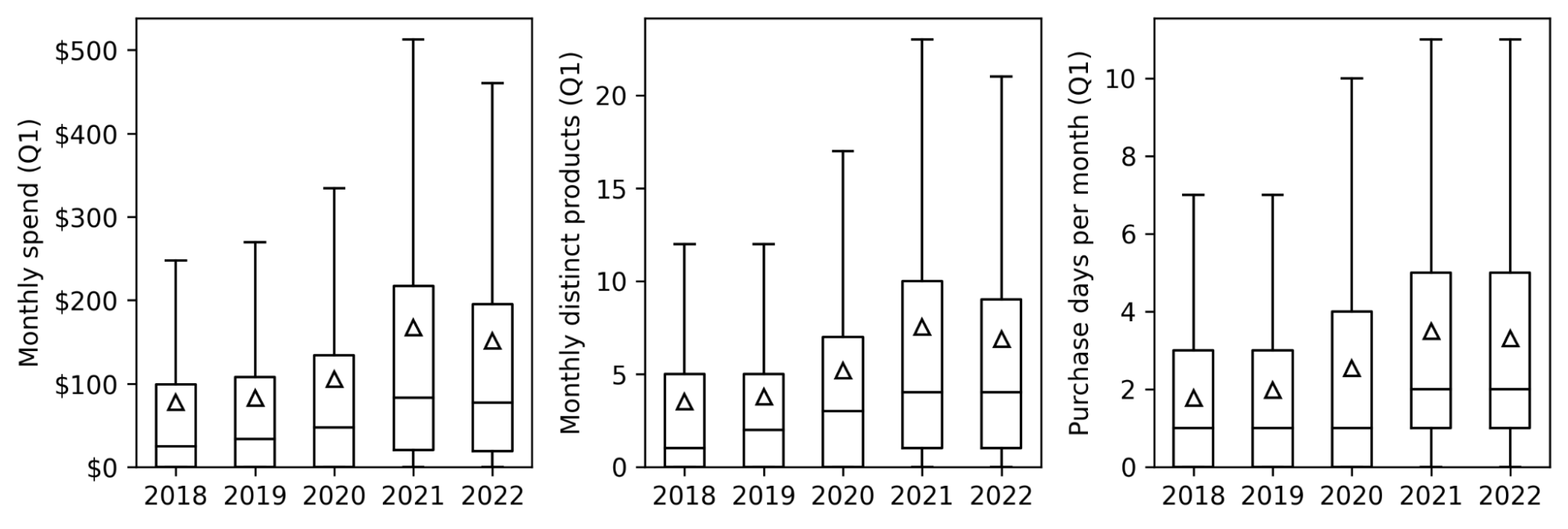}
    \caption{Distribution of monthly metrics across users (n=4115) for Q1 of each year. Boxplots show the medians (lines), means (triangles), first and third quartiles, and whiskers indicate the 1.5 x IQR. Outliers are omitted (see SI Tables~\ref{tab:s6}-\ref{tab:s8}).}
    \label{fig:consumer_metrics_dist}
    \end{adjustwidth}
\end{figure}

This section of the results presents analyses using the purchasing behavior metrics shown in Fig~\ref{fig:amzn_sample_spend_vs_net_sales} computed as panel data: for each user, for each month, we computed total spend, the number of distinct products purchased, and the number of distinct days the user made purchases (purchase frequency). Fig~\ref{fig:consumer_metrics_dist} shows the distribution of user-level monthly metrics averaged over Q1 of each year (median), showing there is large variation in the metrics across individual users. Each metric grew on average by more than 85\% between 2018 and 2022, peaking in 2021 (COVID-19 period).  More details on these distributions are shown in SI Tables~\ref{tab:s6}-\ref{tab:s8} and Fig~\ref{fig:consumer_metrics_dist_detailed}.

\subsubsection{Relationships between spend, distinct products, and purchase days}

The purchasing behavior metrics are highly correlated, as shown in Fig~\ref{fig:amzn_sample_spend_vs_net_sales}. We briefly expand on this relationship and then focus the main analyses on monthly purchase days.

The Pearson correlation between monthly metrics for spend and distinct products is $r=0.757$ ($p<0.001$), for spend and purchase days is $r=0.702$ ($p<0.001$), and for purchase days and distinct products is $r=0.835$ ($p<0.001$). We also estimate a linear relationship between the monthly metrics for distinct products and purchase days. On average, 1 additional purchase day corresponds to approximately 2 additional products purchased each month ($\beta=2.143$; $p<0.001$), with a small but positive interaction effect between time and purchase days ($\beta=0.004; p<0.001$), indicating a positive trend over time. See SI Table~\ref{tab:s9} for details.

\subsubsection{COVID had a limited impact on the trajectory of online purchasing}

\begin{figure}[ht]
\begin{adjustwidth}{-0.6in}{0in}
    \centering
    \includegraphics[width=\linewidth]{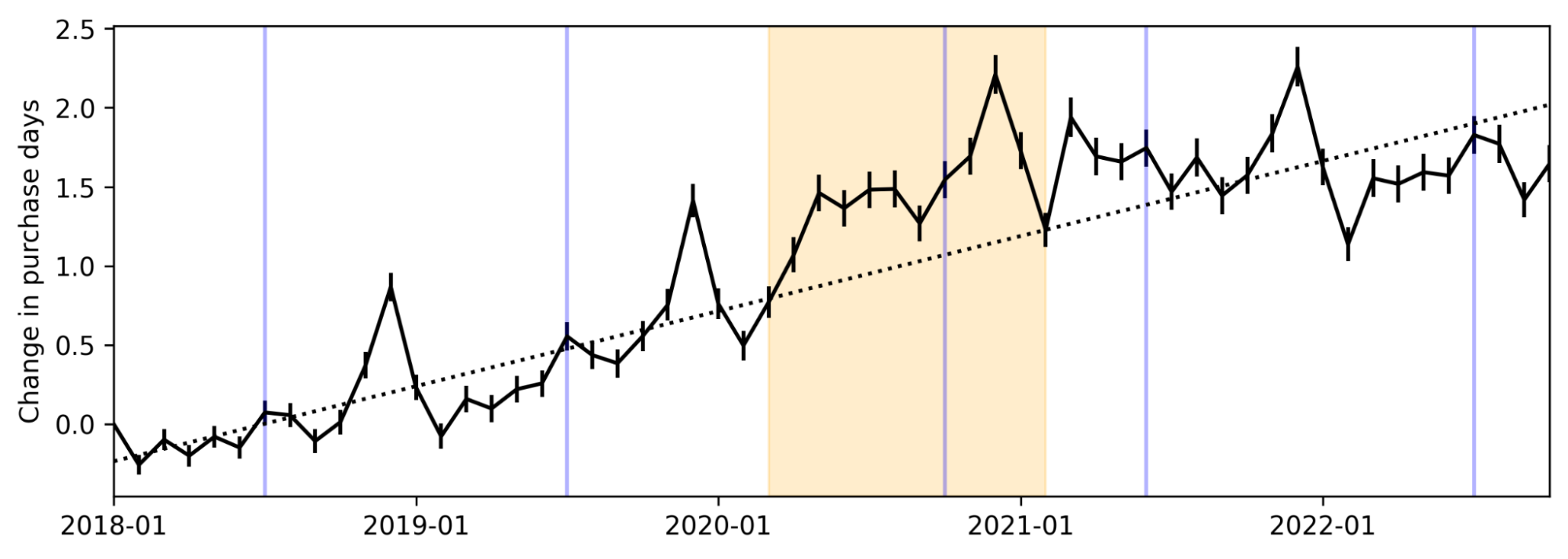}
    \caption{Graphical event study estimating change in purchase frequency (purchase days per month) over time. Solid lines display coefficients with 95\% CIs. The dashed line displays the trend estimated over the pre-pandemic period (2018-01 to 2020-02). The orange section indicates the first year of COVID (2020-03 to 2021-02). Vertical blue lines indicate months Amazon Prime Day occurred.}
    \label{fig:event_study_purchase_days}
    \end{adjustwidth}
\end{figure}

Fig~\ref{fig:event_study_purchase_days} presents a graphical event study which we use to evaluate the impact of COVID on the upward trajectory of consumers' online purchase frequency. It displays the average change in purchase days per month, in a regression that controls for users' sex, age, and income demographics.  Resulting coefficients are shown relative to a 2018-01 baseline. We also estimated the positive trend prior to COVID (dashed line in Fig~\ref{fig:event_study_purchase_days}) via a linear regression trained on the coefficients estimated for 2018-01 to 2020-02 (SI Table~\ref{tab:s10}).

We do not find the COVID-19 pandemic had a lasting impact on the pre-pandemic trend. Instead, we find COVID provided a transient shock, significantly increasing online purchasing behavior above the trend line temporarily ($\beta=0.5578$; $p<0.001$). Purchasing then returned to a level no higher than the pre-pandemic trend would have brought it to. Numerical results are detailed in the SI (Table~\ref{tab:s11}).

For robustness, we repeat this event study using the number of distinct products purchased each month as the dependent variable, instead of purchase days. The results are similar, showing a temporary boost due to COVID, where the metrics then resolve to the pre-pandemic trend line (see SI Fig~\ref{fig:event_study_products}).

\subsubsection{Purchasing behaviors differ by demographic groups}

\begin{figure}[ht]
    \centering
    \includegraphics[width=0.875\linewidth]{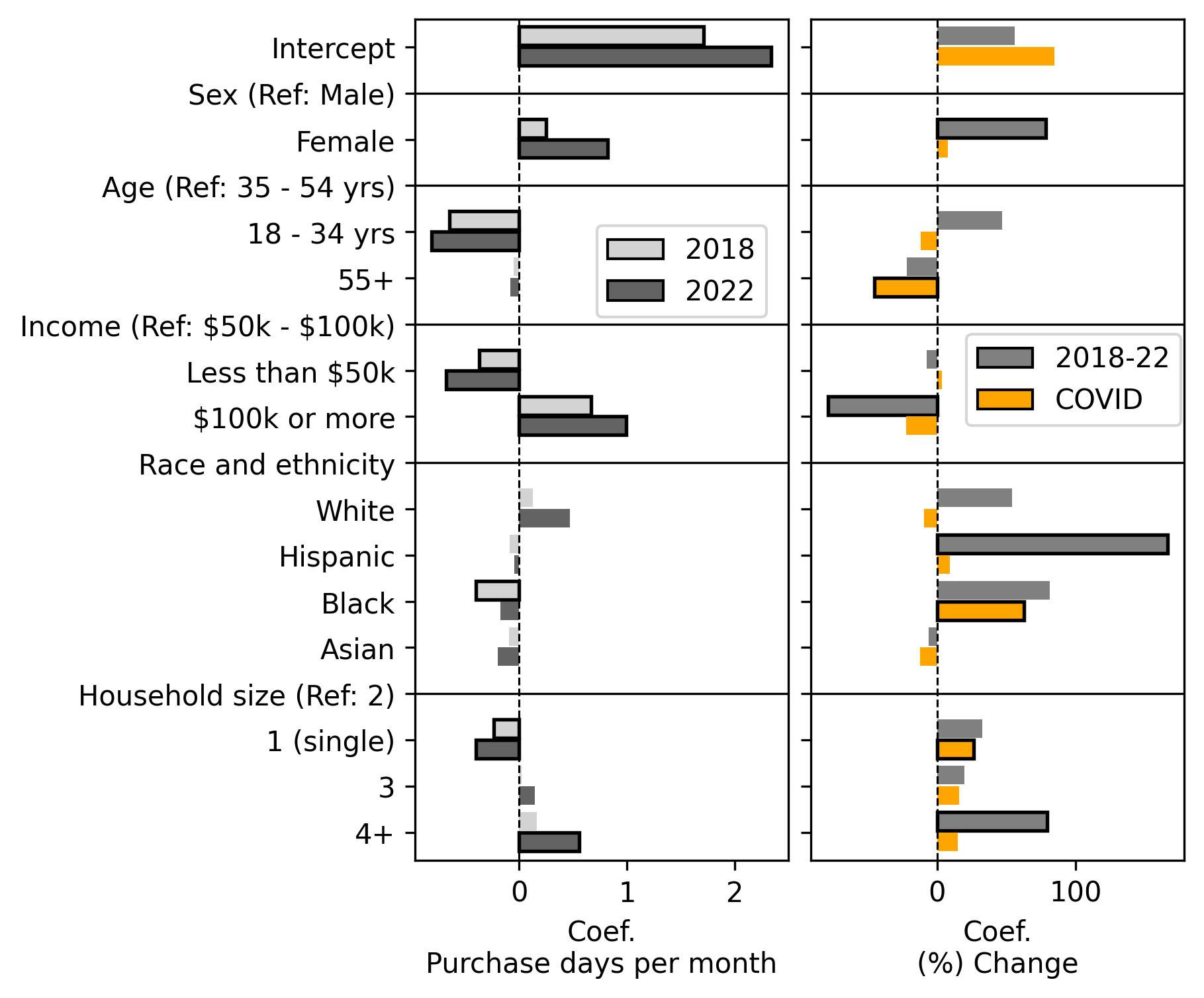}
    \caption{Coefficient estimates reporting relative impact of consumer demographics on (left) purchase frequency for 2018 and 2022, and (right) change in purchase frequency from 2018 to 2022, and from one year prior to COVID to the first year of COVID. Bars indicating statistically significant values (p$<$0.05) are outlined in black.}
    \label{fig:purchase_freq_diffs_by_demos}
\end{figure}

Fig~\ref{fig:purchase_freq_diffs_by_demos} shows results from the regressions analyzing relationships between demographics and purchasing (Eq~\ref{eq:demos_purchase_freq}), where statistically significant values ($p<0.05$) are outlined in black. Numerical results are in SI Table~\ref{tab:s12}.

Four separate OLS regressions were estimated, differing in the dependent variable. The independent variables are the consumers' demographics. For results shown in the left panel of Fig~\ref{fig:purchase_freq_diffs_by_demos}, the dependent variables are consumers' median purchase frequency for 2018 and 2022. The right panel shows the relative differences between demographics when evaluating changes over time. The dependent variables are users' percent change in purchase frequency from 2018 to 2022, and percent change from the year prior to COVID to the first year of COVID.

Results show female consumers made online purchases more frequently than male consumers, and this gap grew from 2018 to 2022. However, we do not find the first year of COVID played a significant role in this growth.

Younger consumers (18 - 34 years) made online purchases significantly less often than middle aged (35 - 54 years) consumers and the oldest consumer group (55 years and older) increased their purchasing behavior less than their middle aged counterparts during the COVID period. And while consumers in the lowest household income group (\$50k or less) purchased the least often, and the highest income group (\$100k or more) purchased the most often, in both 2018 and 2022, purchase frequency for the highest income group grew the least from 2018 to 2022.

Race and ethnicity also played a role, with different purchasing behavior changes seen in the overall 2018 to 2022 study period versus the COVID period. Hispanic consumers increased their purchase frequency from 2018 to 2022, compared to non-Hispanics, but this change was not significant during the COVID period. Black consumers did significantly increase their purchase frequency during COVID, compared to non-Black consumers, and this difference is large. We discuss potential reasons and implications for these differences in the Discussion.
In addition we see an intuitive result where the smallest households purchase least often and the largest households purchase most often. From 2018 to 2022 the largest households (4 or more) saw the greatest increase in purchase frequency. In contrast, in the first year of COVID it was consumers living alone who increased their purchase frequency most.

\subsection{Substitution effects and impacts on brick-and-mortar retail}

While the previous results show changes in consumer purchasing behaviors, this section shows potential consequences for retail, by showing evidence of substitution effects between online and offline retail. 

\subsubsection{Substitution effects for books and shoes prior to COVID}
We tested and validated a hypothesis that there is a negative relationship between online purchases and employment at local retail stores by leveraging yearly state level employment data. We compared year-to-year changes in the number of people making purchases for books and shoes in each state, to the change in employment at book stores and shoe stores in each state. We tested this hypothesis for years 2018 to 2019, prior to COVID, because we expect COVID disrupted how retailers made employment and planning decisions and employment data were not available in later years. 

In this analysis, changes in employment broadly represents changes in availability of local (state) retail options as well as potential growth or contraction of employment within local retailers. We find the expected negative correlation with Pearson $r=-0.344$ ($p=0.021$) and $r=-0.315$ ($p=0.035$) for books and shoes, respectively.

This correlation does not imply a causal relationship in either direction. For example, it is possible more consumers make Amazon purchases for books and shoes when there are fewer local retail options available. It is also possible growth in Amazon purchases leads to reduced revenue and then reduced employment at local retailers. Either way, this negative relationship indicates substitution effects between Amazon and local book and shoe stores.

\subsubsection{Substitution effects demonstrated through COVID}

Again we tested and validated a hypothesis that there is a negative relationship between online and in-store shopping. Here we use data from the COVID-19 pandemic to test for such substitution effects with grocery purchases. The analysis takes advantage of how COVID-19 provided exogenous shocks, and how groceries necessitate frequent recurring purchases, versus other purchases that can more easily be delayed. 

We used mobility data from Google reporting on visits to grocery stores in each state~\cite{googlellc2020}. We compared the store visits data to the number of Amazon users making grocery purchases within each state. We tested the top 3 US states, ordered by the average number of monthly Amazon users purchasing groceries, CA, TX, NY. We find the expected negative correlation with Pearson $r=-0.544$ ($p=0.007$), $r=-0.431$ ($p=0.040$), and $r=-0.513$ ($p=0.012$), for CA, TX, NY, respectively.

\section{Discussion}

E-commerce and COVID-19 have had transformative effects on consumer behavior and retail markets, yet much of the data needed to study these changes remain within Amazon and other private firms.  We begin to address this public knowledge gap by leveraging a unique, recently published dataset of Amazon purchase histories, providing new insights into how online purchasing has changed over time and the impact of COVID-19 on these trends. This work presents a case study in how consumer-level, crowdsourced purchases data have the potential to produce insights beyond those available from common data aggregates.

The itemized and longitudinal nature of the purchases data allowed us to develop new metrics, quantifying changes in consumers' purchase frequency and the diversity of products they purchase online. We argue that compared to expenditure, which prior analyses are limited to, these metrics better quantify changes in consumers' online purchasing behaviors. These consumer-level metrics increased persistently prior to COVID and throughout the pandemic, growing by more than 85\% from 2018 to 2022 on average.  We consider the rapid growth in these purchasing behaviors another kind of ``Amazon effect". Our analyses also show nuanced differences in this growth across consumer demographics and between the overall study period versus the first year of the COVID pandemic. For example, female consumers in the dataset made online purchases significantly more often than their male counterparts, and increased their purchasing significantly more than males from 2018 to 2022, where the first year of COVID did not play a significant role in expanding this gap. The story is different when inspecting other demographic dimensions. The largest households purchased most often, and increased purchasing most from 2018 to 2022 compared to other household sizes, which may be intuitive. Yet single person households increased their purchasing most during the first year of COVID. Differences by race demonstrate how these differences may be important. Black consumers significantly increased their purchase frequency over non-Black consumers during COVID, and this difference is large. When considering why, we note that Black Americans suffered larger losses due to the pandemic~\cite{aburto2022} - they were overall hospitalized at 2x the rate of White Americans~\cite{cdc2023} - and early analyses by major news outlets brought attention to this racial disparity~\cite{eligon2020}. If this led to Black consumers adapting their shopping behaviors to reduce COVID risk factors, this may be reflected in their substantial increase in online purchase frequency during the pandemic. These results present an example of how disaggregated purchases data can help shed light on how economic and health related shocks may have differing impacts on various consumer groups.

Our analyses also fill in research gaps left from the pandemic period, where researchers and analysts projected COVID would accelerate e-commerce adoption, with the long-term impacts unknown~\cite{pantano2020,ratchford2022,roggeveen2020}. We find COVID had a limited impact on the trajectory of the purchasing behaviors we studied. While our results show COVID significantly increased purchasing behaviors initially, we also find the metrics in 2022 were no greater than the result of the pre-pandemic trend. 

The Amazon purchases also provide evidence of substitution effects between online and offline retail for the books, shoes, and grocery sectors. While our analyses do not present a causal relationship, previous researchers have, showing how the rollout of Amazon fulfillment centers reduced sales and employment for nearby retail establishments~\cite{chava2023}. The researchers described these changes as e-commerce driven ``creative destruction", as they also found corresponding employment gains in the transportation-warehousing and food services sectors, although these gains did not outweigh the losses for retail. Our analyses add to the larger discussion of Amazon's effect on local retail by showing how substitution effects occur in specific retail sectors where Amazon has made substantial investments. However, the limitations of our dataset limit our analyses to the state level and a small window of time. More local and causal analyses should be pursued to better understand these potential ``Amazon effects".

\subsection{Limitations and future work}
These results are limited due to the limited timespan of the purchases data.  It is possible purchasing behaviors plateaued, or it is possible that past 2022 they continued along the pre-pandemic trajectory. US Census Bureau data that reports quarterly estimates of e-commerce retail sales~\cite{ECOMNSA2024} suggests e-commerce again trended upwards in the months following our study period (see SI Fig~\ref{fig:ECOMNSA_vs_sample_spend}). However, these census data track sales rather than purchasing behaviors, without controlling for prices and inflation, and they aggregate sales from a variety of undisclosed e-commerce firms, versus just Amazon. Further research would benefit from more crowdsourced Amazon purchases data.

In addition to timespan, our analyses are limited by the dataset's geographic granularity and user sample.  For example, we use US states to evaluate and control for geographic differences, yet there is likely heterogeneity within states or across urban, suburban and rural areas. More work should be done to understand the role of geography in the transition from offline to online purchasing.  With respect to the sample, the sampled users were already making purchases by 2018, which may leave gaps in the demographics analyses. Future work should explore how newer Amazon users engage with online versus offline retail. In particular, there will be a cohort of young consumers who came of age entering digital economies during COVID-19 lockdowns and their purchasing behaviors may be very different. 

Furthermore, more robust and detailed analyses are limited by the dataset's sample size, which is small relative to Amazon's user base, which includes the majority of American consumers~\cite{npr/maristpoll2018}. Even with these limitations, this work demonstrates how economic insights can be gained through open data crowdsourced from consumers, providing initial analyses for future researchers to build upon with more data.

\section*{Data and code availability}
All data and code used in this analysis are available via the following open repository: \url{https://github.com/aberke/amazon-study}.

\bibliography{references}

\begin{thebibliography}{10}

\bibitem{censusEcommerce}
{US Census Bureau}. Monthly {{Retail Trade}} - {{Quarterly Retail E-Commerce Sales Report}};.
\newblock Available from: \url{https://www.census.gov/retail/ecommerce.html}.

\bibitem{emarketerRanks2023}
Droesch B. Target, {{Carvana}}, and {{Lowe}}’s Are Moving up the {{US}} Ecommerce Sales Rank. EMARKETER; 2023.
\newblock Available from: \url{https://www.emarketer.com/content/target-carvana-lowe-s-moving-up-us-ecommerce-sales-ranks}.

\bibitem{mamaghani2020}
Mamaghani EJ, Davari S.
\newblock The Bi-Objective Periodic Closed Loop Network Design Problem.
\newblock Expert Systems with Applications. 2020;144:113068.
\newblock Available from: \url{https://www.sciencedirect.com/science/article/pii/S0957417419307857}.

\bibitem{daugherty2019}
Daugherty PJ, Bolumole Y, Grawe SJ.
\newblock The New Age of Customer Impatience.
\newblock International Journal of Physical Distribution \& Logistics Management. 2019;49(1):4-32.
\newblock Available from: \url{https://doi.org/10.1108/IJPDLM-03-2018-0143}.

\bibitem{AmazonEffectsInvestopedia}
Mitchell C. Amazon {{Effect}}: {{Definition}}, {{Statistics}}, {{Impact}} on {{Consumers}}. Investopedia; 2024.
\newblock Available from: \url{https://www.investopedia.com/terms/a/amazon-effect.asp}.

\bibitem{ourSciDataPaper}
Berke A, Calacci D, Mahari R, Yabe T, Larson K, Pentland S.
\newblock Open E-Commerce 1.0, Five Years of Crowdsourced {{U}}.{{S}}. {{Amazon}} Purchase Histories with User Demographics.
\newblock Nature Scientific Data. 2024;11(1):491.
\newblock Available from: \url{https://www.nature.com/articles/s41597-024-03329-6}.

\bibitem{chen2016}
Chen L, Mislove A, Wilson C.
\newblock An {{Empirical Analysis}} of {{Algorithmic Pricing}} on {{Amazon Marketplace}}.
\newblock In: Proceedings of the 25th {{International Conference}} on {{World Wide Web}}. International World Wide Web Conferences Steering Committee; 2016. p. 1339-49.
\newblock Available from: \url{https://dl.acm.org/doi/10.1145/2872427.2883089}.

\bibitem{cavallo2018}
Cavallo A. More {{Amazon Effects}}: {{Online Competition}} and {{Pricing Behaviors}}. National Bureau of Economic Research; 2018.
\newblock Available from: \url{http://www.nber.org/papers/w25138.pdf}.

\bibitem{ater2018}
Ater I, Rigbi O. The {{Effects}} of {{Mandatory Disclosure}} of {{Supermarket Prices}} [SSRN Scholarly Paper]; 2018.
\newblock Available from: \url{https://dx.doi.org/10.2139/ssrn.3178561}.

\bibitem{chava2023}
Chava S, Oettl A, Singh M, Zeng L.
\newblock Creative {{Destruction}}? {{Impact}} of {{E-Commerce}} on the {{Retail Sector}}.
\newblock Management Science. 2023;70(4):2168-87.
\newblock Available from: \url{https://doi.org/10.1287/mnsc.2023.4795}.

\bibitem{ghose2006}
Ghose A, Smith MD, Telang R.
\newblock Internet {{Exchanges}} for {{Used Books}}: {{An Empirical Analysis}} of {{Product Cannibalization}} and {{Welfare Impact}}.
\newblock Information Systems Research. 2006;17(1):3-19.
\newblock Available from: \url{https://pubsonline.informs.org/doi/10.1287/isre.1050.0072}.

\bibitem{forman2009}
Forman C, Ghose A, Goldfarb A.
\newblock Competition {{Between Local}} and {{Electronic Markets}}: {{How}} the {{Benefit}} of {{Buying Online Depends}} on {{Where You Live}}.
\newblock Management Science. 2009;55(1):47-57.
\newblock Available from: \url{https://pubsonline.informs.org/doi/10.1287/mnsc.1080.0932}.

\bibitem{amazon1998}
Amazon. Amazon.Com {{Acquires Three Leading Internet Companies}}; 1998.
\newblock Available from: \url{https://press.aboutamazon.com/1998/4/amazon-com-acquires-three-leading-internet-companies}.

\bibitem{zapposSEC2009}
SEC. Press {{Release}}: {{Amazon}}.Com to {{Acquire Zappos}}.Com; 2009.
\newblock Available from: \url{https://www.sec.gov/Archives/edgar/data/1018724/000119312509153130/dex991.htm}.

\bibitem{sec2017}
SEC. {{FORM}} 8-{{K}} | {{Whole Foods Market}}, {{Inc}}. United States Securities and Exchange Commission; 2017.
\newblock Available from: \url{https://www.sec.gov/Archives/edgar/data/865436/000114420417045261/v474128_8k.htm}.

\bibitem{pantano2020}
Pantano E, Pizzi G, Scarpi D, Dennis C.
\newblock Competing during a Pandemic? {{Retailers}}’ Ups and Downs during the {{COVID-19}} Outbreak.
\newblock Journal of Business Research. 2020;116:209-13.
\newblock Available from: \url{https://www.sciencedirect.com/science/article/pii/S0148296320303209}.

\bibitem{ratchford2022}
Ratchford B, Soysal G, Zentner A, Gauri DK.
\newblock Online and Offline Retailing: {{What}} We Know and Directions for Future Research.
\newblock Journal of Retailing. 2022;98(1):152-77.
\newblock Available from: \url{https://www.sciencedirect.com/science/article/pii/S0022435922000070}.

\bibitem{roggeveen2020}
Roggeveen AL, Sethuraman R.
\newblock How the {{COVID-19 Pandemic May Change}} the {{World}} of {{Retailing}}.
\newblock Journal of Retailing. 2020;96(2):169-71.
\newblock Available from: \url{https://www.sciencedirect.com/science/article/pii/S0022435920300208}.

\bibitem{censusbureauPandemicEconImpact}
Roman S, Cooke-Hull S, Dunfee M, Flaherty M, Haskell J, Holland V, et~al.
\newblock The {{Coronavirus Pandemic}}'s {{Economic Impact}}.
\newblock US Census Bureau Library.
\newblock Available from: \url{https://www.census.gov/library/publications/2022/econ/coronavirus-pandemics-economic-impact.html}.

\bibitem{ourCSCWpaper}
Berke A, Mahari R, Pentland S, Larson K, Calacci D.
\newblock Insights from an Experiment Crowdsourcing Data from Thousands of {{US Amazon}} Users: {{The}} Importance of Transparency, Money, and Data Use.
\newblock Proceedings of the ACM Human-Computer Interaction. 2024;CSCW2:48.
\newblock Available from: \url{https://doi.org/10.1145/3687005}.

\bibitem{AmazonIRQuarterly}
Amazon. Amazon.Com, {{Inc}}. - {{Quarterly}} Results; 2023.
\newblock Available from: \url{https://ir.aboutamazon.com/quarterly-results/default.aspx}.

\bibitem{ECOMNSA2024}
{US Census Bureau}. E-{{Commerce Retail Sales}} [{{ECOMNSA}}]; 2024.
\newblock Available from: \url{https://fred.stlouisfed.org/series/ECOMNSA}.

\bibitem{amazonstaff2019}
{Amazon Staff}. The History of {{Prime Day}}: {{A}} Look Back at {{Amazon}}'s Biggest Deal Event of the Summer. Amazon.com; 2019.
\newblock Available from: \url{https://www.aboutamazon.com/news/retail/the-history-of-prime-day}.

\bibitem{SUSB}
{US Census Bureau}. Statistics of {{U}}.{{S}}. {{Businesses Tables}};.
\newblock Available from: \url{https://www.census.gov/programs-surveys/susb/data/tables.html}.

\bibitem{censusPop2021}
{US Census Bureau, Population Division}. Annual {{Estimates}} of the {{Resident Population}} for the {{United States}}, {{Regions}}, {{States}}, the {{District}} of {{Columbia}}, and {{Puerto Rico}}: {{April}} 1, 2010 to {{July}} 1, 2019; {{April}} 1, 2020; and {{July}} 1, 2020 ({{NST-EST2020}});.

\bibitem{berkeMobilePhonesCovid2024}
Berke A, Larson K.
\newblock Chapter 3 - {{Mobile}} Phones and Their Use to Study Dynamics of the {{COVID-19}} Pandemic.
\newblock In: Rajendram R, Preedy VR, Patel VB, editors. Features, {{Transmission}}, {{Detection}}, and {{Case Studies}} in {{COVID-19}}. Academic Press; 2024. p. 25-37.
\newblock Available from: \url{https://doi.org/10.1016/B978-0-323-95646-8.00049-4}.

\bibitem{googlellc2020}
{Google LLC}. {{COVID-19 Community Mobility Report}}; 2020.
\newblock Available from: \url{https://www.google.com/covid19/mobility?hl=en}.

\bibitem{censusPopStates2022}
{US Census Bureau, Population Division}. Estimates of the {{Total Resident Population}} and {{Resident Population Age}} 18 {{Years}} and {{Older}} for the {{United States}}, {{Regions}}, {{States}}, {{District}} of {{Columbia}}, and {{Puerto Rico}}: {{July}} 1, 2022 ({{SCPRC-EST2022-18}}+{{POP}});.
\newblock Available from: \url{https://www.census.gov/data/tables/time-series/demo/popest/2020s-national-detail.html}.

\bibitem{aburto2022}
Aburto JM, Tilstra AM, Floridi G, Dowd JB.
\newblock Significant impacts of the {COVID}-19 pandemic on race/ethnic differences in {US} mortality.
\newblock Proceedings of the National Academy of Sciences. 2022;119(35):e2205813119.
\newblock Available from: \url{https://www.pnas.org/doi/abs/10.1073/pnas.2205813119}.

\bibitem{cdc2023}
{CDC}. Risk for {COVID}-19 Infection, Hospitalization, and Death By Race/Ethnicity; 2023.
\newblock Available from: \url{https://archive.cdc.gov/www_cdc_gov/coronavirus/2019-ncov/covid-data/investigations-discovery/hospitalization-death-by-race-ethnicity.html}.

\bibitem{eligon2020}
Eligon J, Burch ADS, Searcey D, Jr RAO.
\newblock Black Americans Face Alarming Rates of Coronavirus Infection in Some States.
\newblock The New York Times. 2020.
\newblock Available from: \url{https://www.nytimes.com/2020/04/07/us/coronavirus-race.html}.

\bibitem{npr/maristpoll2018}
{NPR/Marist Poll}. The {{Digital Economy}}: {{Profiles}} of {{Online Consumers}} and {{Survey Methods}}; 2018.
\newblock Available from: \url{https://maristpoll.marist.edu/wp-content/misc/usapolls/us180423_NPR/NPR_Marist%20Poll_Summary%20of%20the%20Profiles%20and%20Profile%20Tables_May%202018.pdf}.

\bibitem{u.s.censusbureaupopulationdivisionDP05ACSDemographicHousing2022}
{US Census Bureau, Population Division}. {{DP05ACS Demographic}} and {{Housing Estimates}};.
\newblock Available from: \url{https://data.census.gov/table?q=United+States+Race+and+Ethnicity&g=&tid=ACSDP1Y2021.DP05}.

\bibitem{u.s.censusbureaupopulationdivisionNCEST2022AGESEXRESAnnualEstimates2023}
{US Census Bureau, Population Division}. {{NC-EST2022-AGESEX-RES}}: {{Annual Estimates}} of the {{Resident Population}} by {{Single Year}} of {{Age}} and {{Sex}} for the {{United States}}: {{April}} 1, 2020 to {{July}} 1, 2022;.

\bibitem{USCensus-HINC-01}
{US Census Bureau, Current Population Survey}. 2023 Annual Social and Economic Supplement ({CPS} {ASEC}). Selected Characteristics of Households by Total Money Income in 2022. Table {HINC}-01.;.

\bibitem{nytimesZappos2009}
Stone B.
\newblock Amazon’s {{Expanding With Deal}} for {{Zappos}}.
\newblock The New York Times. 2009.
\newblock Available from: \url{https://www.nytimes.com/2009/07/23/technology/companies/23amazon.html}.

\bibitem{stevens2017}
Stevens L, Gasparro A.
\newblock Amazon to {{Buy Whole Foods}} for \$13.7 {{Billion}}.
\newblock Wall Street Journal. 2017.
\newblock Available from: \url{https://www.wsj.com/articles/amazon-to-buy-whole-foods-for-13-7-billion-1497618446}.

\end{thebibliography}

\bibliographystyle{vancouver}

\clearpage

\setcounter{figure}{0}
\renewcommand{\thefigure}{S\arabic{figure}}
\setcounter{table}{0}
\renewcommand{\thetable}{S\arabic{table}}

\section*{Supporting Information}

\subsection*{Purchases data example rows}
Each row in the purchases dataset represents a purchase by a particular user and includes an order date, product code (ASIN/ISBN), product title, per-unit price and quantity, state the item was shipped to, and a category assigned by Amazon. Purchases are linked to a single user and their demographics data via a response ID. To better illustrate the dataset, we provide an example set of rows in Table~\ref{tab:s1}.

\begin{table}[!htp]
\begin{adjustwidth}{-2.4in}{0in}
\centering
\caption{Example rows from one user's Amazon purchases data.}
\label{tab:s1}
\scriptsize
\begin{tabular}{llllp{5.7cm}llll}
\toprule
\textbf{Date} &\textbf{Price} &\textbf{Qty} &\textbf{State} &\textbf{Title} &\textbf{ASIN/ISBN} &\textbf{Category} &\textbf{ResponseID} \\
\midrule
2018-01-21 &\$23.07 &1 &OK &OTTERBOX SYMMETRY SERIES Case for iPhone 8 PLUS &B01K6PBRSW &CELLULAR\_PHONE\_CASE &R\_2zARigFd \\
2018-02-06 &\$15.91 &1 &OK &Strength in Stillness: The Power of Transcendental Meditation &1501161210 &ABIS\_BOOK &R\_2zARigFd \\
2018-04-03 &\$5.99 &1 &OK &Square Reader for magstripe (with headset jack) &B00HZYK3CO &MEMORY\_CARD\_READER &R\_2zARigFd \\
2018-06-11 &\$4.89 &1 &OK &Dove Advanced Care Antiperspirant Deodorant Stick for Women &B00Q70R41U &BODY\_DEODORANT &R\_2zARigFd \\
\bottomrule
\end{tabular}
\end{adjustwidth}
\end{table}

\subsection*{Sample demographics}

Tables~\ref{tab:s1}-\ref{tab:s3} report on the Amazon user sample demographics. We include comparisons to the US population using US census data for sex~\cite{u.s.censusbureaupopulationdivisionDP05ACSDemographicHousing2022}, age~\cite{u.s.censusbureaupopulationdivisionNCEST2022AGESEXRESAnnualEstimates2023}, and household income~\cite{USCensus-HINC-01}. Given the users in the Amazon sample are at least 18 years of age, we compare the sample data to census data for the 18 or older population. Users responded about their race and ethnicity via multiple choice; counts indicate whether a category was selected at all and are not expected to sum to the total N=5027.

\begin{table}[!htp]
\caption{Sample demographics compared to US census data.}
\label{tab:s2}
\begin{tabular}{lrrrr}\toprule
&\multicolumn{2}{c}{\textbf{Survey}} &\textbf{Census} \\\cmidrule{2-4}
\textbf{Attribute} &N &\% &\% \\\midrule
\textbf{Gender} & & & \\
Female &2589 &51.5\% &51\% \\
Male &2311 &46.0\% &49\% \\
Other &127 &2.5\% & \\
\textbf{Age} & & & \\
18 - 34 years &2581 &51.4\% &29.4\% \\
35 - 54 years &1917 &38.1\% &32.3\% \\
55 and older &529 &10.5\% &38.3\% \\
\textbf{Household income} & & & \\
Less than \$50,000 &1874 &37.3\% &35.5\% \\
\$50,000 - \$99,999 &1824 &36.3\% &30.3\% \\
\$100,000 or more &1253 &24.9\% &34.1\% \\
Prefer not to say &76 &1.5\% & \\
\bottomrule
\end{tabular}
\end{table}

\begin{table}[!ht]
\caption{Sample race and ethnicity.}
\label{tab:s3}
\begin{tabular}{lrr}\toprule
\textbf{Race / ethnicity} &\textbf{N} \\
\midrule
White &4133 \\
Hispanic &549 \\
Asian &483 \\
Black &448 \\
Other &263 \\
\bottomrule
\end{tabular}
\end{table}

\begin{table}[!ht]
\caption{Sample household sizes.}
\label{tab:s4}
\begin{tabular}{lrr}\toprule
\textbf{Household size} &\textbf{N} \\
\midrule
1 &1199 \\
2 &1590 \\
3 &983 \\
4+ &1255 \\
\bottomrule
\end{tabular}
\end{table}

\subsection*{E-commerce retail sales and sample expenditure}

\begin{figure}[ht]
\begin{adjustwidth}{-1.7in}{0in}
    \centering
    \includegraphics[width=\linewidth]{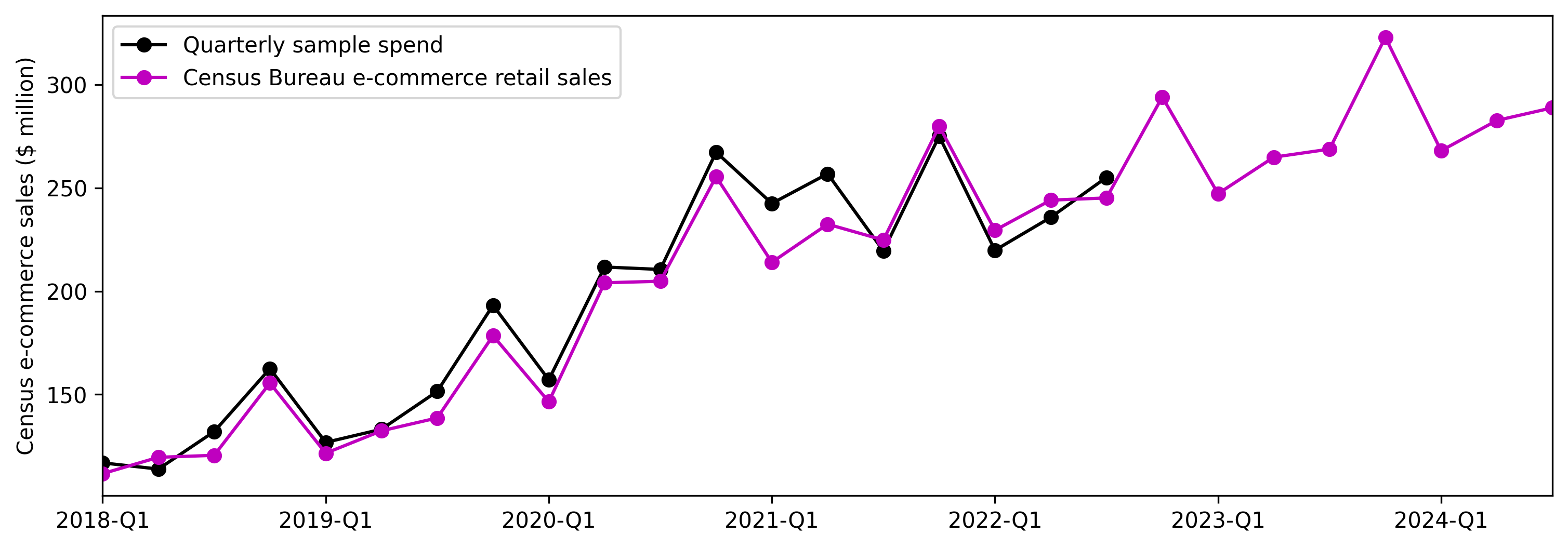}
    \caption{Quarterly e-commerce retail sales from the Census Bureau compared to the quarterly expenditure from our sample data, extending beyond our study period.}
    \label{fig:ECOMNSA_vs_sample_spend}
    \end{adjustwidth}
\end{figure}

Fig~\ref{fig:ECOMNSA_vs_sample_spend} shows quarterly e-commerce retail sales from the US Census Bureau~\cite{ECOMNSA2024} from our study period extending to 2024-Q3, which is the most recently available data at the time of analysis. For comparison, we show our sample data, also aggregated to quarterly expenditure. The Pearson correlation between the sample and census expenditure data computed over our study period (2018-Q1 to 2022-Q3), is $r=0.982$ ($p<0.001$).

\clearpage
\subsection*{Robustness check on the role of groceries in the sample metrics}

\begin{figure}[!ht]
\begin{adjustwidth}{-1.7in}{0in}
    \begin{flushright}
    \includegraphics[width=\linewidth]{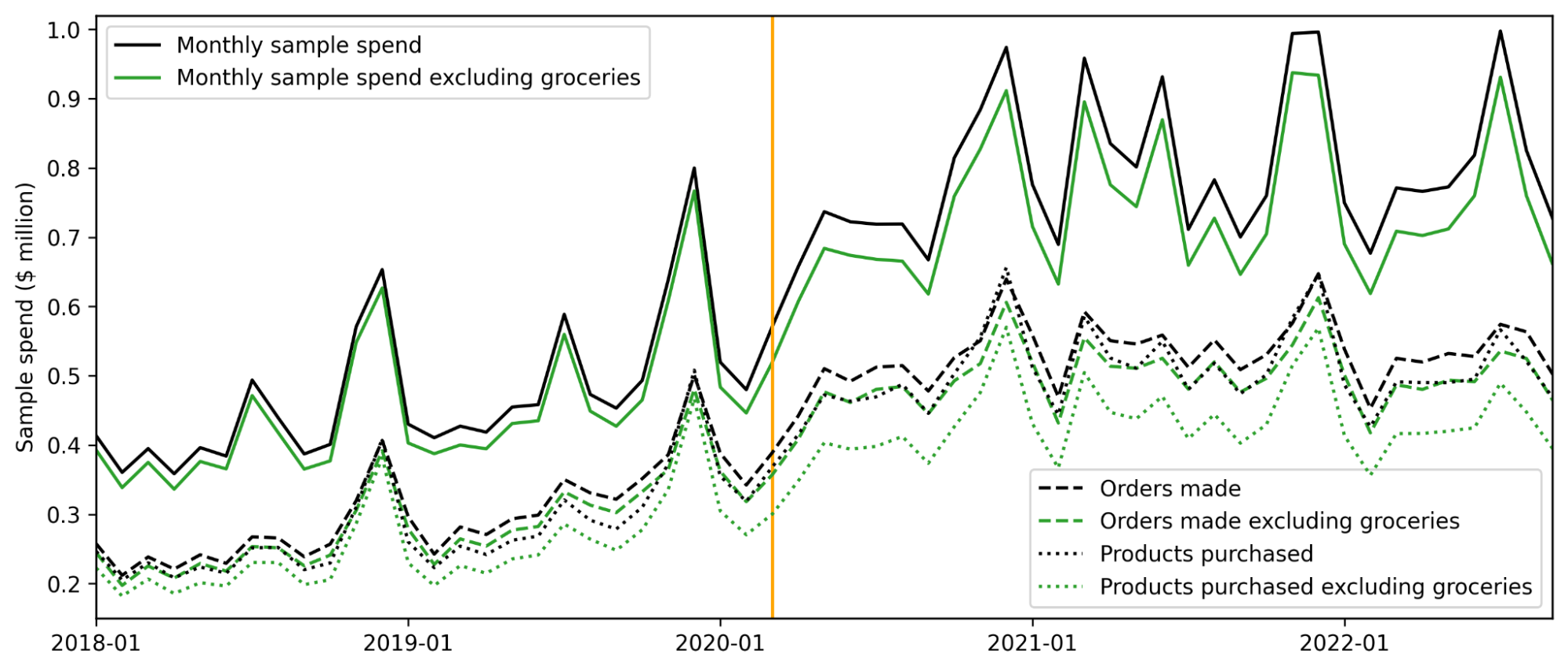}
    \caption{Purchasing metrics computed with and without grocery products.}
    \label{fig:amzn_sample_spend_vs_net_sales_without_grocery}
    \end{flushright}
\end{adjustwidth}
\end{figure}

For robustness, we check whether the patterns in the purchasing behavior metrics are consistent with and without the inclusion of grocery purchases. We reproduce the monthly spend, products, and frequency time series shown in Fig~\ref{fig:amzn_sample_spend_vs_net_sales} with and without grocery related purchases, and plot the metrics in Fig~\ref{fig:amzn_sample_spend_vs_net_sales_without_grocery}.  We can see groceries helped drive growth in these metrics, yet the metrics are still highly correlated, showing that the patterns are relatively consistent across the purchases data with and without the grocery purchases. For sample spend: Pearson $r=0.999$ ($p<0.001$); for orders made: $r=0.999$ ($p<0.001$); for products purchased: $r=0.996$ ($p<0.001$).

\subsection*{The impact of Amazon Prime Day on purchase behavior metrics}

Amazon's major recurring sales event, ``Prime Day", occurred during the following months in our data (24): 2018-07, 2019-07, 2020-10, 2021-06, 2022-07. We statistically test for the impact of Prime Day on purchase behaviors for the users in our panel data using the following OLS model: \newline

\begin{math}
    y_{i,t} = a_i + \beta_1 \cdot isPrimeMonth_t  + \beta_2 \cdot t +  \sum_m \beta_m \cdot month_t + \beta \cdot X_i + e_i	
\end{math} \newline

Where $y_{i,t}$ is the purchasing behavior metric (distinct products/purchase days) for month $t$ by user $i$. $isPrimeMonth_t$ is the variable of interest, set to 1 when the month includes Prime Day, 0 otherwise. Results are reported in Table~\ref{tab:s5}, showing Prime Day significantly increased monthly purchasing metrics ($p<0.001$).

\begin{table}[!ht]
\begin{threeparttable}
\caption{OLS regression results estimating the impact of Amazon Prime Day on monthly purchasing metrics.}\label{tab:s5}
\begin{tabular}{lrrr}
\toprule
&\textbf{Purchase days} &\textbf{Distinct products} \\
\midrule
\textbf{Intercept} &2.035*** (0.093) &3.832*** (0.229) \\
\textbf{Sex (Ref: Male)} & & \\
Female &0.586*** (0.083) &1.810*** (0.209) \\
\textbf{Age (Ref: 35 - 54 yrs)} & & \\
18 - 34 yrs &-0.881*** (0.090) &-2.198*** (0.232) \\
55+ &-0.202 (0.140) &-0.946** (0.344) \\
\textbf{Income (Ref: \$50k - \$100k)} & & \\
Less than \$50k &-0.581*** (0.089) &-1.196*** (0.225) \\
\$100k or more &0.954*** (0.115) &2.240*** (0.292) \\
\textbf{Month (Ref: 1)} & & \\
2 &-0.404*** (0.018) &-0.808*** (0.057) \\
3 &-0.083*** (0.020) &0.019 (0.069) \\
4 &-0.151*** (0.020) &-0.230*** (0.063) \\
5 &-0.055** (0.020) &-0.136* (0.063) \\
6 &-0.141*** (0.021) &-0.251*** (0.065) \\
7 &-0.125*** (0.024) &-0.304*** (0.079) \\
8 &-0.055** (0.020) &-0.087 (0.064) \\
9 &-0.299*** (0.020) &-0.710*** (0.062) \\
10 &-0.188*** (0.021) &-0.459*** (0.067) \\
11 &0.139*** (0.023) &0.861*** (0.076) \\
12 &0.624*** (0.024) &2.175*** (0.086) \\
\textbf{t} &0.039*** (0.001) &0.094*** (0.003) \\
\textbf{is Prime Day Month} &0.171*** (0.019) &0.733*** (0.063) \\
\midrule
\textbf{N} &229738 &229738 \\
\textbf{R-squared} &0.087 &0.067 \\
\bottomrule
\end{tabular}
\begin{tablenotes}
  \small
  \item Significance denoted as *p$<$0.05; **p$<$0.01; ***p$<$0.001.
\end{tablenotes}
\end{threeparttable}
\end{table}

\clearpage

\subsection*{Distribution of purchase behavior metrics and changes over time}

\begin{figure}[!ht]
\begin{adjustwidth}{-0.55in}{0in}
    \centering
    \includegraphics[width=\linewidth]{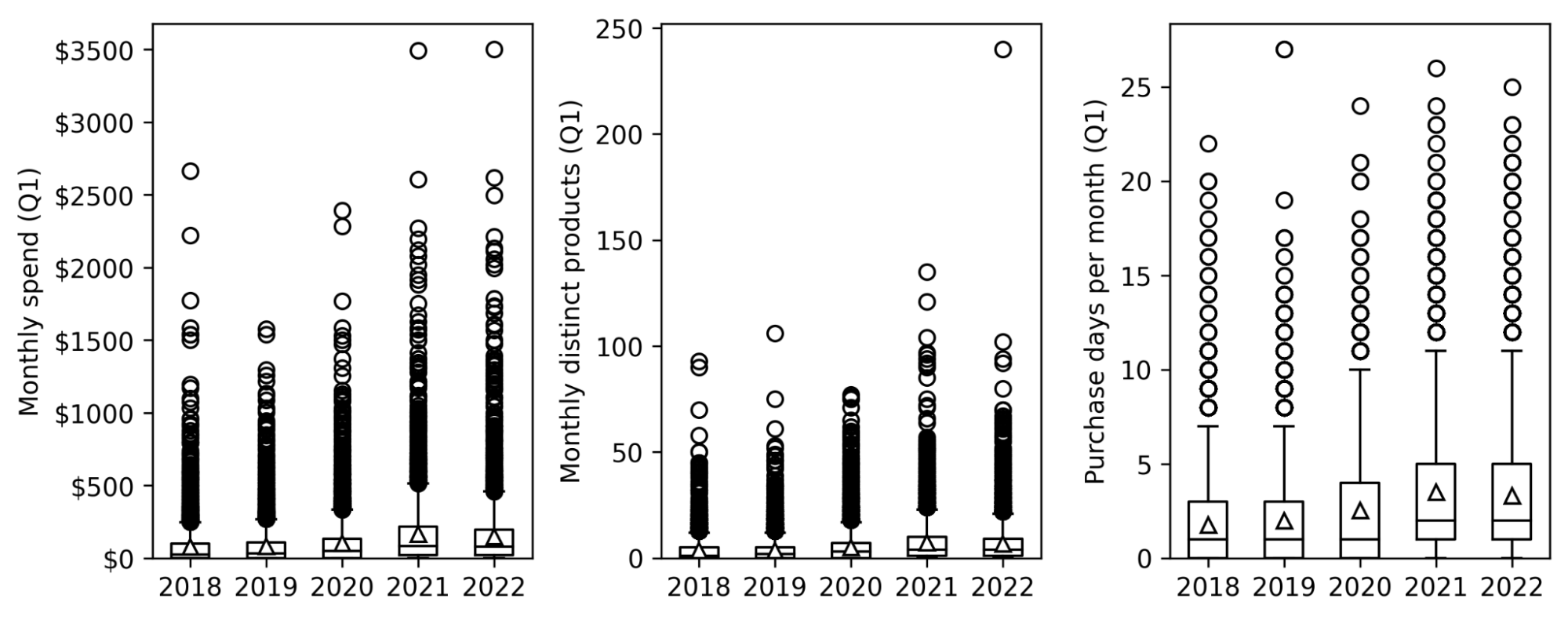}
    \caption{Fig \ref{fig:consumer_metrics_dist} reproduced to include outliers. The boxplots show the distribution of monthly metrics across users (n=4115).}
    \label{fig:consumer_metrics_dist_detailed}
    \end{adjustwidth}
\end{figure}

\begin{table}[ht]
\caption{Distribution of monthly total spend per user averaged over Q1 (n=4115).}
\label{tab:s6}
\begin{tabular}{lrrrrrr}
\toprule
\textbf{Year} &\textbf{2018} &\textbf{2019} &\textbf{2020} &\textbf{2021} &\textbf{2022} \\\midrule
\textbf{Mean} &77.49 &83.05 &105.46 &167.15 &151.67 \\
\textbf{Std} &147.56 &136.55 &168.91 &241.57 &229.89 \\
\textbf{Min} &0 &0 &0 &0 &0 \\
\textbf{25\%} &0 &0 &0 &19.95 &18.97 \\
\textbf{50\%} &24.99 &33.55 &47.59 &83.09 &76.89 \\
\textbf{75\%} &99.08 &108 &133.98 &217.1 &195.41 \\
\textbf{Max} &2665.77 &1575.2 &2391.92 &3492.65 &3501.27 \\
\bottomrule
\end{tabular}
\end{table}

\begin{table}[!htp]
\caption{Distribution of monthly distinct products purchased per user averaged over Q1 (n=4115).}
\label{tab:s7}
\begin{tabular}{lrrrrrr}
\toprule
\textbf{Year} &\textbf{2018} &\textbf{2019} &\textbf{2020} &\textbf{2021} &\textbf{2022} \\\midrule
\textbf{Mean} &3.54 &3.77 &5.19 &7.54 &6.9 \\
\textbf{Std} &5.93 &5.95 &7.86 &10.42 &10.08 \\
\textbf{Min} &0 &0 &0 &0 &0 \\
\textbf{25\%} &0 &0 &0 &1 &1 \\
\textbf{50\%} &1 &2 &3 &4 &4 \\
\textbf{75\%} &5 &5 &7 &10 &9 \\
\textbf{Max} &93 &106 &77 &135 &240 \\
\bottomrule
\end{tabular}
\end{table}

\begin{table}[!htp]
\caption{Distribution of monthly purchase days per user averaged over Q1 (n=4115).}
\label{tab:s8}
\begin{tabular}{lrrrrrr}
\toprule
\textbf{Year} &\textbf{2018} &\textbf{2019} &\textbf{2020} &\textbf{2021} &\textbf{2022} \\
\midrule
\textbf{Mean} &1.78 &1.98 &2.55 &3.5 &3.31 \\
\textbf{Std} &2.43 &2.53 &3.07 &3.75 &3.61 \\
\textbf{Min} &0 &0 &0 &0 &0 \\
\textbf{25\%} &0 &0 &0 &1 &1 \\
\textbf{50\%} &1 &1 &1 &2 &2 \\
\textbf{75\%} &3 &3 &4 &5 &5 \\
\textbf{Max} &22 &27 &24 &26 &25 \\
\bottomrule
\end{tabular}
\end{table}

\subsection*{Relationship between purchase behavior metrics}

To estimate the relationship between distinct products and purchase days each month, we estimate the following OLS regression using the panel data: \newline

\begin{math}
distinctProducts_{i,t} = \beta_1 \cdot purchaseDays_{i,t} +  \beta_2 \cdot t \cdot purchaseDays_{i,t}
\end{math} \newline
			
Results are shown in Table~\ref{tab:s9}.

\begin{table}[!htp]
\caption{Regression results estimating relationship between distinct products purchased and purchase days per month.}
\label{tab:s9}
\begin{threeparttable}
\begin{tabular}{rll}
\toprule
\textbf{} &\textbf{Coef} \\
\midrule
\textbf{Purchase days} &2.1428*** (0.006) \\
\textbf{Purchase days x t} &0.0037*** (0.000) \\
\midrule
\textbf{Observations} &238670 \\
\textbf{R-squared} &0.785 \\
\bottomrule
\end{tabular}
\begin{tablenotes}
  \small
  \item *p$<$0.05; **p$<$0.01; ***p$<$0.001.
\end{tablenotes}
\end{threeparttable}
\end{table}

\subsection*{Event study with consumer purchasing behavior metrics}

\begin{figure}[ht]
\begin{adjustwidth}{-0.6in}{0in}
    \centering
    \includegraphics[width=\linewidth]{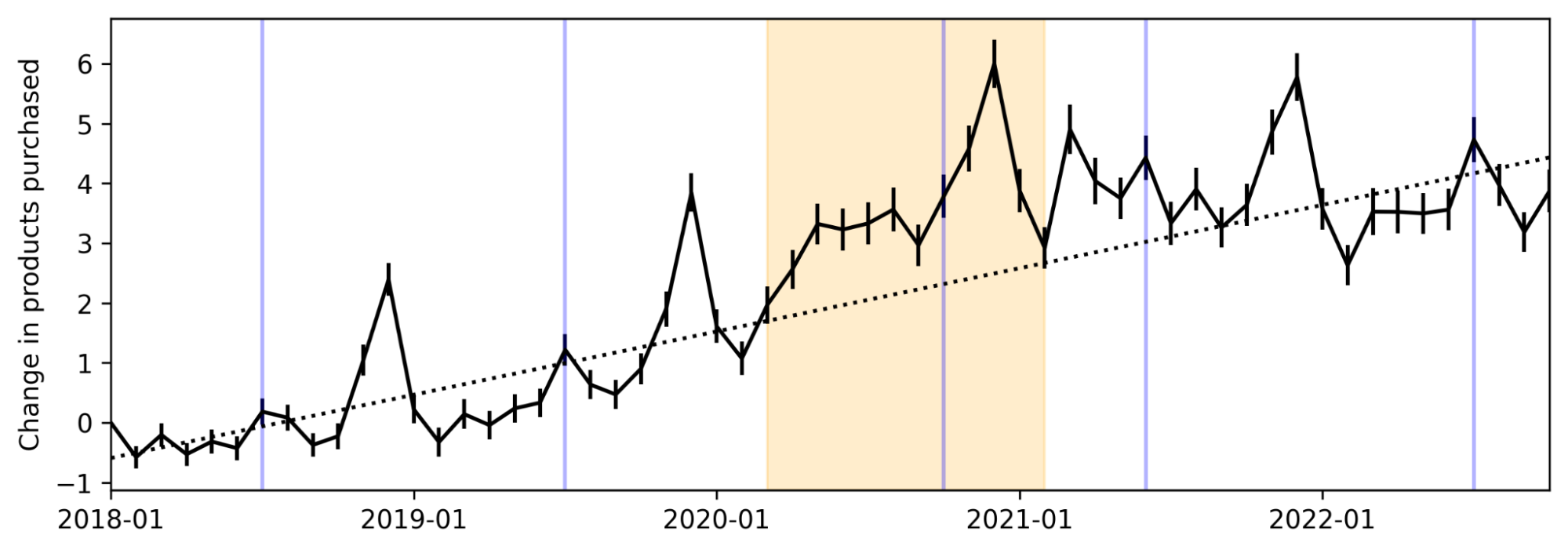}
    \caption{Graphical event study estimating change in distinct products purchased over time, using the same analysis methods as Fig \ref{fig:event_study_purchase_days}.}
    \label{fig:event_study_products}
    \end{adjustwidth}
\end{figure}

\begin{figure}[ht]
\begin{adjustwidth}{-0.6in}{0in}
    \centering
    \includegraphics[width=\linewidth]{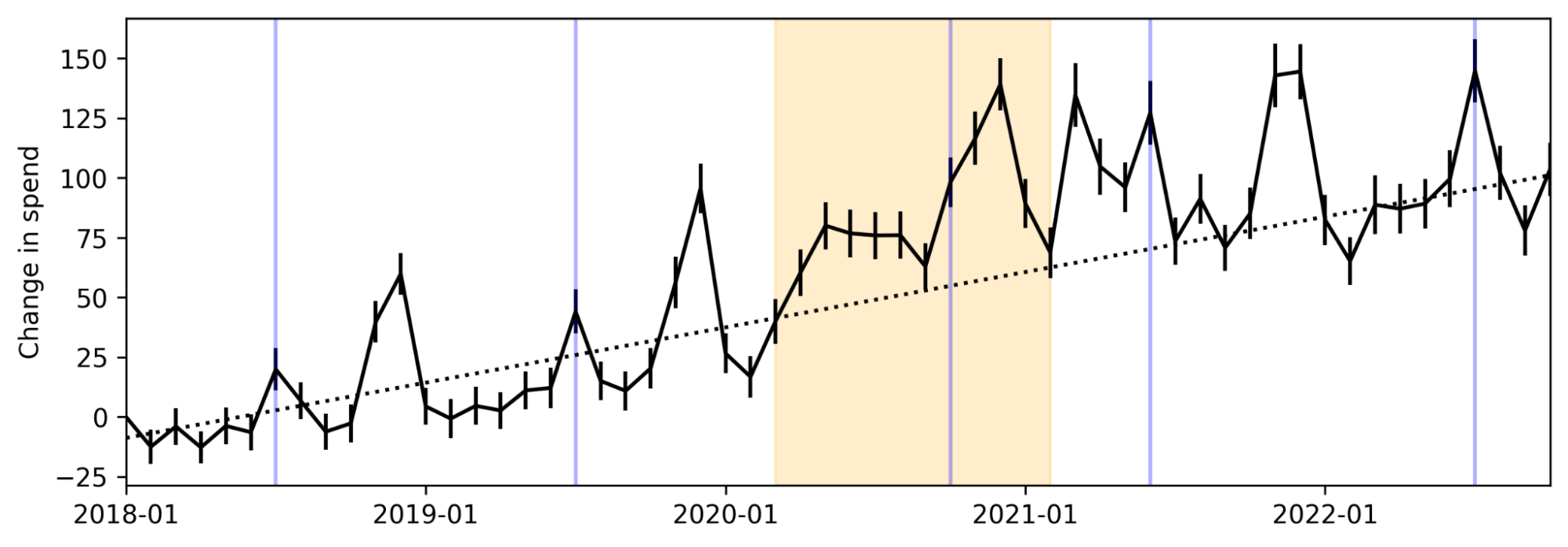}
    \caption{Graphical event study estimating change in spend (\$USD) over time, using the same analysis methods as Fig \ref{fig:event_study_purchase_days}.}
    \label{fig:event_study_spend}
    \end{adjustwidth}
\end{figure}

\begin{table}[ht]
\caption{Regression results for Eq~\ref{eq:graphical_event_study_2} estimating trend for graphical event study shown in Fig~\ref{fig:event_study_purchase_days}.}
\label{tab:s10}
\begin{threeparttable}
\begin{tabular}{rll}
\toprule
&\textbf{Coef} \\
\midrule
\textbf{Intercept} &-0.2362* (0.098) \\
\textbf{t} &0.0396*** (0.007) \\
\midrule
\textbf{Observations} &26 \\
\textbf{R-squared} &0.591 \\
\bottomrule
\end{tabular}
\begin{tablenotes}
  \small
  \item *p$<$0.05; **p$<$0.01; ***p$<$0.001.
\end{tablenotes}
\end{threeparttable}
\end{table}

\begin{table}[!htp]
\caption{Event study regression results for Eq~\ref{eq:event_study_3}. The positive significant term for $postCOVID$ indicates the additional purchase days in the $postCOVID$ period are significantly higher than the trend.}
\label{tab:s11}
\begin{threeparttable}
\begin{tabular}{rll}\toprule
\textbf{Variable} &\textbf{Coef} \\
\midrule
Intercept &2.0878 (0.089) \\
\textbf{Month (ref: 1)} & \\
2 &-0.3667*** (0.019) \\
3 &-0.3206*** (0.024) \\
4 &-0.3071*** (0.024) \\
5 &-0.1284*** (0.023) \\
6 &-0.2043*** (0.023) \\
7 &-0.0245 (0.024) \\
8 &-0.1007*** (0.024) \\
9 &-0.2780*** (0.022) \\
10 &-0.1209*** (0.023) \\
11 &0.0819** (0.025) \\
12 &0.6069*** (0.026) \\
\textbf{Sex (ref: male)} & \\
Female &0.4574*** (0.077) \\
\textbf{Age (ref: 35 - 54 years)} & \\
18 - 34 years &-0.8400*** (0.084) \\
55 years and older &-0.1442 (0.131) \\
Income (ref \$50,000 - \$99,999) & \\
\$100,000 or more &0.9046*** (0.107) \\
Less than \$50,000 &-0.5181*** (0.082) \\
\textbf{t} &0.0326*** (0.002) \\
\textbf{Post COVID} &0.5578*** (0.035) \\
\midrule
\textbf{Observations} &150518 \\
\textbf{R-squared} &0.089 \\
\bottomrule
\end{tabular}
\begin{tablenotes}
  \small
  \item Significance denoted as *p$<$0.05; **p$<$0.01; ***p$<$0.001.
\end{tablenotes}
\end{threeparttable}
\end{table}

\subsection*{Relationships between demographics and purchasing behavior}

\begin{figure}[ht]
\begin{adjustwidth}{-2.25in}{0in}
    \centering
    \includegraphics[width=\linewidth]{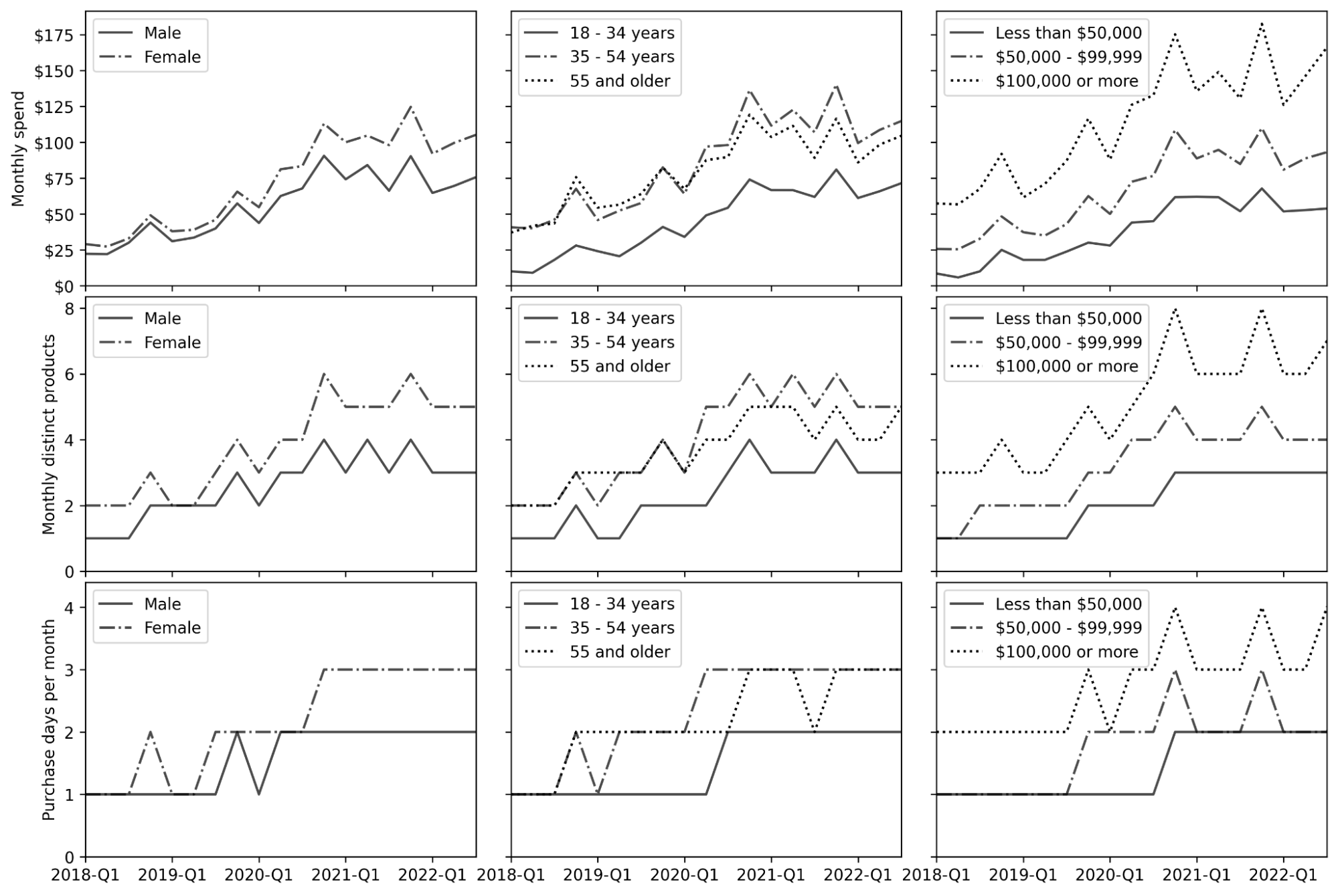}
    \caption{Variation in consumer level metrics across demographic groups, with lines showing the median computed across users in each group.}
    \label{fig:consumer_metrics_by_demos}
    \end{adjustwidth}
\end{figure}

Fig~\ref{fig:consumer_metrics_by_demos} shows an overview of trends in purchasing behavior across demographics over time. For each of the three purchasing behavior metrics, it shows the median over the consumer-level metrics for each demographic group. 
This shows how female consumers in our sample spent more, bought more different products, and made purchases more often than males, and this consumption gap expanded over the study period. Younger consumers (18 - 34 years) spent less, bought fewer distinct products, and made purchases less often than older consumers. Fig~\ref{fig:consumer_metrics_by_demos} also compares purchasing behaviors across users' household income levels. As might be expected, spending increased with household income, with the highest income group (\$100k or more) spending most and the lowest income group (\$50k or less) spending least, on average. However, Fig~\ref{fig:consumer_metrics_by_demos} shows these expenditure differences are largely consistent with the purchasing behavior metrics (distinct products and purchase frequency), demonstrating how higher income earners' higher spending may be partly due to them buying more items more often, rather than just buying more expensive items. Fig~\ref{fig:consumer_metrics_by_demos} also shows differences in holiday (Q4) spending patterns. In particular, higher and medium income households show sharper spikes in purchasing metrics versus lower income households.
A limitation of these plots is that they evaluate one category of consumer demographics at a time (e.g. age) whereas these categories may be related (e.g. younger consumers may tend to have lower household incomes). The regression analysis shown in Fig~\ref{fig:purchase_freq_diffs_by_demos} (via Eq~\ref{eq:demos_purchase_freq}) controls for these demographics and their relationships.

\subsubsection*{Regression analysis}
Table~\ref{tab:s12} shows regression results for Eq~\ref{eq:demos_purchase_freq}, which evaluate the relationships between consumers' demographics and purchase frequency. 	
Four separate OLS models were estimated, only differing in the dependent variable: (1) median monthly purchase days for 2018, (2) median monthly purchase days in 2022, (3) percent change in purchase days from 2018 to 2022, (3) percent change in purchase days from the year prior to COVID (2019-03 to 2020-02) to the period spanning the first year of COVID-19 (2020-03 to 2021-02). In addition to the demographic variables shown in Fig~\ref{fig:purchase_freq_diffs_by_demos}, the OLS models included fixed effects for each users' US state of residence.

\begin{table}[ht]
\begin{adjustwidth}{-1.79in}{0in} 
\centering
\caption{Regression results estimating relationships between demographics and purchase frequency.}
\label{tab:s12}
\begin{threeparttable}
\begin{tabular}{rlllll}
\toprule
&\textbf{(1)} &\textbf{(2)} &\textbf{(3)} &\textbf{(4)} \\\midrule
Intercept &1.710*** (0.346) &2.338*** (0.525) &56.154 (128.563) &84.468 (47.150) \\
\textbf{Sex (Ref: Male)} & & & & \\
Female &0.250*** (0.071) &0.819*** (0.108) &78.612** (26.239) &7.628 (9.517) \\
\textbf{Age (Ref: 35 - 54 yrs)} & & & & \\
18 - 34 yrs &-0.647*** (0.077) &-0.807*** (0.117) &47.216 (28.652) &-11.636 (10.366) \\
55+ &-0.049 (0.119) &-0.082 (0.181) &-21.914 (43.626) &-44.968** (15.930) \\
\textbf{Income (Ref: \$50k - \$100k)} & & & & \\
Less than \$50k &-0.367*** (0.085) &-0.677*** (0.129) &-7.296 (31.374) &3.402 (11.353) \\
\$100k or more &0.665*** (0.091) &0.995*** (0.138) &-78.699* (33.417) &-22.133 (12.125) \\
\textbf{Race and ethnicity} & & & & \\
White &0.129 (0.162) &0.468 (0.246) &54.350 (60.959) &-9.470 (21.922) \\
Hispanic &-0.089 (0.126) &-0.045 (0.191) &166.143*** (46.673) &9.195 (17.001) \\
Black &-0.397* (0.179) &-0.170 (0.272) &81.351 (67.585) &62.943** (24.171) \\
Asian &-0.092 (0.179) &-0.200 (0.271) &-6.167 (67.155) &-12.318 (24.193) \\
\textbf{Household size (Ref: 2)} & & & & \\
1 (single) &-0.232* (0.098) &-0.398** (0.149) &32.877 (36.079) &26.368* (13.090) \\
3 &0.026 (0.102) &0.145 (0.155) &20.017 (37.646) &16.108 (13.642) \\
4+ &0.164 (0.097) &0.554*** (0.147) &79.425* (35.760) &15.126 (12.938) \\
\textbf{State F.E.} &Yes &Yes &Yes &Yes \\
\midrule
\textbf{N} &3961 &3961 &3792 &3862 \\
\textbf{R-squared} &0.087 &0.109 &0.023 &0.02 \\
\bottomrule
\end{tabular}
\begin{tablenotes}
  \small
  \item Significance denoted as *p$<$0.05; **p$<$0.01; ***p$<$0.001.
\end{tablenotes}
\end{threeparttable}
\end{adjustwidth}
\end{table}

\subsection*{Amazon acquisitions and product categories}

When analyzing relationships between Amazon purchases and brick-and-mortar retail, we focus on three retail sectors in which Amazon has substantially invested: Books, shoes, and grocery. Amazon started as an online marketplace for books, describing itself as ``Earth's Biggest Bookstore", and acquired other book sellers to extend its catalog~\cite{amazon1998}.  In 2009, Amazon acquired the leading online footwear company Zappos for \$1.2 billion~\cite{zapposSEC2009}, which was the company's largest acquisition at the time~\cite{nytimesZappos2009}. In 2017, Amazon purchased the grocery chain Whole Foods Market Inc. for \$13.7 billion and analysts speculated the grocery stores could help Amazon expand its distribution network~\cite{stevens2017,sec2017}. 

\subsubsection*{Product categories}
We use the `Category' label that Amazon assigned to each purchase in our dataset to determine whether a purchase is categorized in our analyses as `Books', `Shoes', or `Grocery'. 

\paragraph{Books purchases.} 
Categories used: `ABIS\_BOOK', `BOOK', `BOOKS\_1973\_AND\_LATER'.

We analyze a total of 83,140 book purchases from N=4,188 users.

\paragraph{Shoes purchases.} 
Categories used: `SHOES', `TECHNICAL\_SPORT\_SHOE', `BOOT', `SANDAL', `SLIPPER'.

We analyze a total of 20,885 shoes purchases from N=3,344 users.

\paragraph{Grocery purchases.} 
Categories used: `GROCERY', `FOOD', `VEGETABLE', `FRUIT', `DAIRY\_BASED\_CHEESE', `BREAD', `POULTRY', `DRINK\_FLAVORED', `SNACK\_CHIP\_AND\_CRISP', `HERB', `MEAT', `FRUIT\_SNACK', `SAUCE', `NUT\_AND\_SEED', `SNACK\_MIX', `WATER', `DAIRY\_BASED\_CREAM', `NOODLE', `PUFFED\_SNACK', `PACKAGED\_SOUP\_AND\_STEW', `MILK\_SUBSTITUTE', `CRACKER', `COOKIE', `SUGAR\_CANDY', `SYRUP', `DAIRY\_BASED\_BUTTER', `BREAKFAST\_CEREAL', `COFFEE', `TEA', `SNACK\_FOOD\_BAR', `POPCORN', `LEAVENING\_AGENT', `Grocery', `DAIRY\_BASED\_YOGURT', `CHOCOLATE\_CANDY', `EDIBLE\_OIL\_VEGETABLE', `JUICE\_AND\_JUICE\_DRINK', `SEASONING', `CAKE', `DAIRY\_BASED\_ICE\_CREAM', `DAIRY\_BASED\_DRINK', `CONDIMENT', `LEGUME', `RICE\_MIX', `CHEWING\_GUM', `FISH', `CULINARY\_SALT', `SUGAR', `NUTRITIONAL\_SUPPLEMENT', `NUT\_BUTTER', `PASTRY', `THICKENING\_AGENT', `CEREAL', `BAKING\_MIX', `SALAD\_DRESSING', `HERBAL\_SUPPLEMENT'.

We analyze a total of 219,749 grocery purchases from N=4,185 users.

\subsection*{Google mobility reports and Amazon grocery purchases}

\begin{figure}[ht]
\begin{adjustwidth}{-0.25in}{0in}
    \centering
    \includegraphics[width=\linewidth]{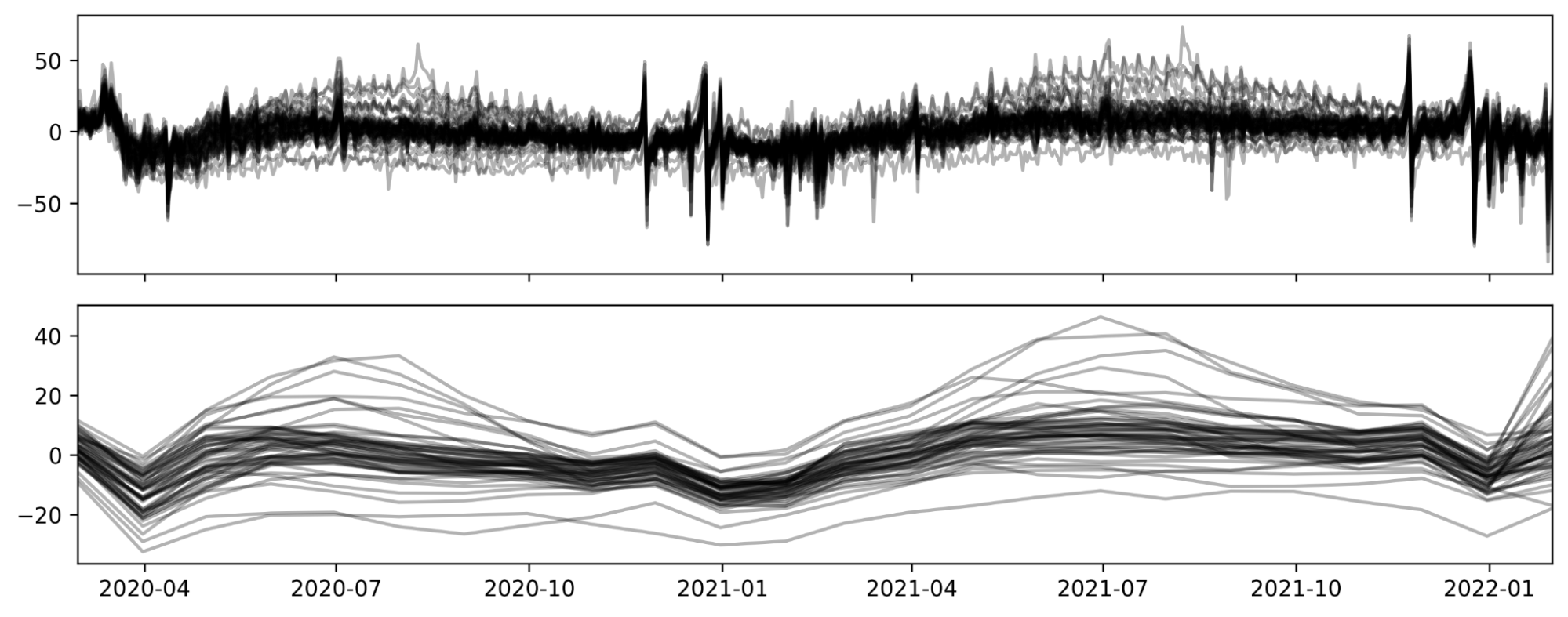}
    \caption{Google COVID-19 mobility reports ``Grocery and pharmacy" indices, for each US state, (top) reported daily and (bottom) aggregated to monthly means.}
    \label{fig:google_covid_mobility}
    \end{adjustwidth}
\end{figure}

\begin{figure}[ht]
\begin{adjustwidth}{-0.25in}{0in}
    \centering
    \includegraphics[width=\linewidth]{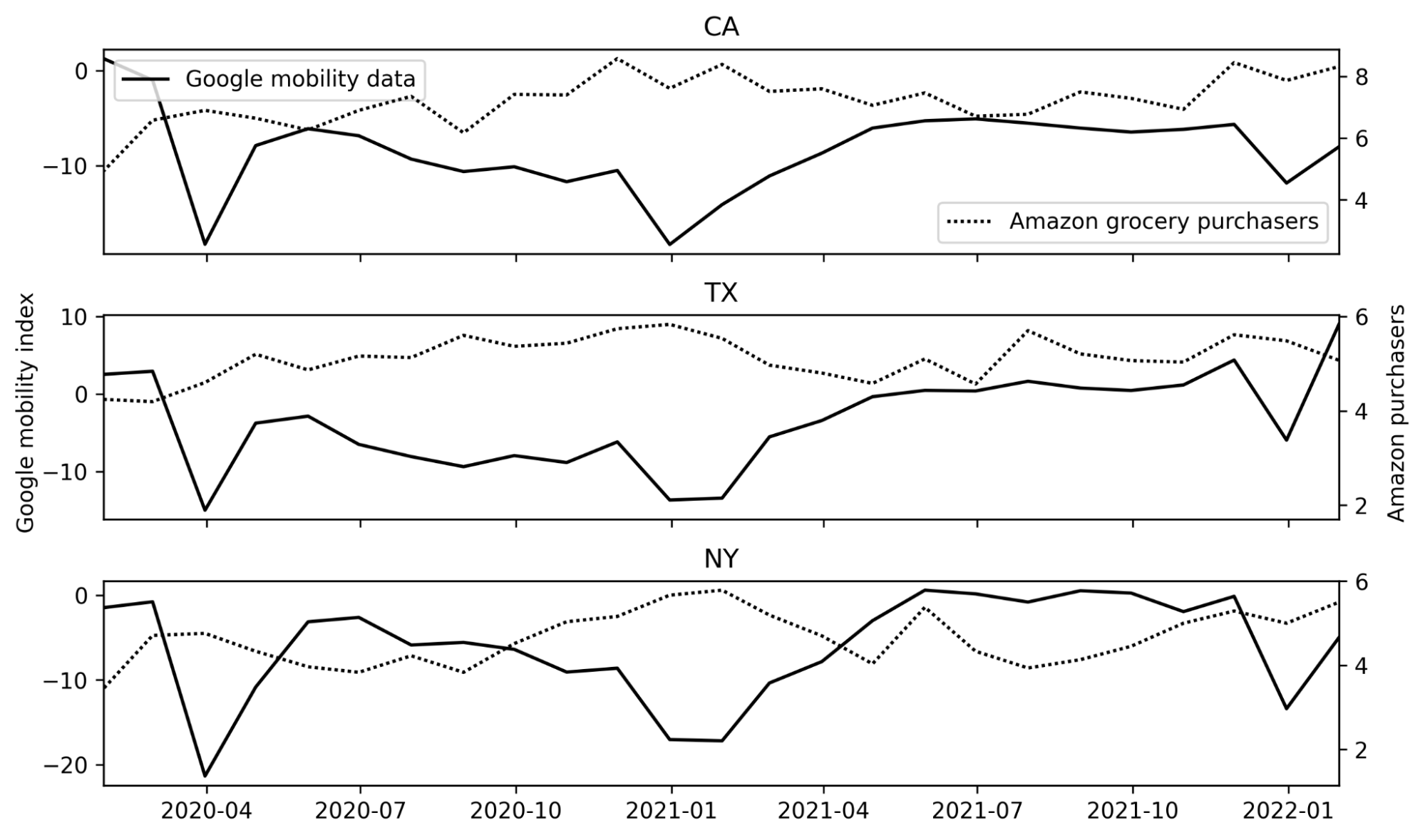}
    \caption{Monthly averages for the Google COVID-19 mobility reports ``Grocery and pharmacy" indices compared to the monthly number of Amazon users making purchases for groceries for CA (r=-0.544; p=0.007), TX (r=-0.431; p=0.040) and NY (r=-0.513; p=0.012).}
    \label{fig:google_covid_mobility_vs_purchases}
    \end{adjustwidth}
\end{figure}

\end{document}